\documentclass[fleqn,usenatbib,useAMS]{mnras}
\usepackage{graphicx}	
\usepackage{amsmath}	
\usepackage{amssymb}	
\usepackage{multicol}
\usepackage{bm}		
\usepackage{pdflscape}	
\usepackage{tabularx}
\usepackage{subcaption}
\usepackage{xfrac}
\usepackage{placeins}

\usepackage[T1]{fontenc}
\usepackage{ae,aecompl}

\usepackage{newtxtext,newtxmath}

\title[Detection of Quadruple Structure Near the ASCC\,32 Region]{Detection of Quadruple Structure Near the ASCC\,32 Region via Machine Learning Methods}

\author[M. Noormohammadi]{
M. Noormohammadi,$^{1}$ \thanks{E-mail: monoorastro@ipm.ir}
A. Javadi,$^{1}$\thanks{E-mail: atefeh@ipm.ir}
and M. Khakian Ghomi$^{2}$
\\
$^{1}$School of Astronomy, Institute for Research in Fundamental Sciences (IPM), P.O.Box 1956836613, Tehran, IRAN\\
$^{2}$Physics and Energy Engineering Department, Amirkabir University, Tehran, IRAN\\
}

\date{Accepted XXX. Received YYY; in original form ZZZ}

\pubyear{2025}

\begin{document}
\label{firstpage}
\pagerange{\pageref{firstpage}--\pageref{lastpage}}
\maketitle
\color{black}

\begin{abstract}
\noindent
Multiple structures within stellar groups are an intriguing subject for theoretical and observational studies of stellar formation. With the accuracy and completeness of data from Gaia Data Release 3, we now have new opportunities to detect reliable members of stellar groups across a larger field of view than in previous studies. In this work, using machine learning methods and high-accuracy data, we investigate the possibility of detecting multiple structures within 500 arcmin of ASCC\,32. We first applied DBSCAN to proper motion and parallax, as multiple structures tend to share similar values for these parameters. Next, we applied GMM to position, proper motion, and parallax for the members detected by DBSCAN. This approach allowed us to identify a filamentary structure among the DBSCAN-detected members. This structure contains all stellar groups previously identified in this region. Subsequently, based on the BIC score, we applied GMM to this filamentary structure. Since multiple structures exhibit distinct positional distributions, GMM was able to effectively separate all groups within the filament. Our methods successfully identified ASCC\,32, OC\,0395, and HSC\,1865 within a 500 arcmin radius. Additionally, we found two distinct substructures within ASCC\,32. These four groups exhibit a single main-sequence distribution in the CMD, with proper motion values within three times the standard deviation and slightly differing parallax values, despite having distinct spatial structures. Furthermore, these four groups share the same radial velocity distribution. We provide documentation demonstrating the formation of these stellar groups as a multiple structure, with improved membership identification compared to previous studies.

\end{abstract}


\section{Introduction}
\noindent
\begin{table*}
\centering
\caption{GDR3 astrometric data uncertainties}
\begin{tabular}{ccccc}
    \hline
    \hline
    Name & G$<$15~(mag) & G$=$17~(mag) & G$=$20~(mag) & G$=$21~(mag) \\
    \hline
    Position& $0.01-0.02$\,mas & $0.05$\,mas & $0.4$\,mas & $1.0$\,mas  \\
    Proper Motion & $0.02-0.03$\,mas\,yr$^{-1}$ & $0.07$\,mas\,yr$^{-1}$ & $0.5$\,mas\,yr$^{-1}$ & $1.4$\,mas\,yr$^{-1}$\\
    Parallax & $0.02-0.03$\,mas & $0.07$\,mas & $0.5$\,mas & $1.3$\,mas\\
    \hline
  \end{tabular}
  \label{uncertainties_astro.tab}
\end{table*}
  \begin{table*}
\centering
\caption{GDR3 photometric data uncertainties}
\begin{tabular}{cccc}
    \hline
    \hline
    Name & G$<$13~(mag) & G$=$17~(mag) & G=20~(mag) \\
    \hline
   G band & $0.3$\,mmag &$1$\,mmag &$6$\,mmag \\
    G$_{BP}$ band & $0.9$\,mmag &$12$\,mmag &$108$\,mmag\\
    G$_{RP}$ band & $0.6$\,mmag &$6$\,mmag &$52$\,mmag\\
    \hline
  \end{tabular}
  \label{uncertainties_photo.tab}
\end{table*}
\begin{figure*}
    \centering
    \begin{subfigure}{0.48\textwidth}
        \centering
        \includegraphics[width=\textwidth]{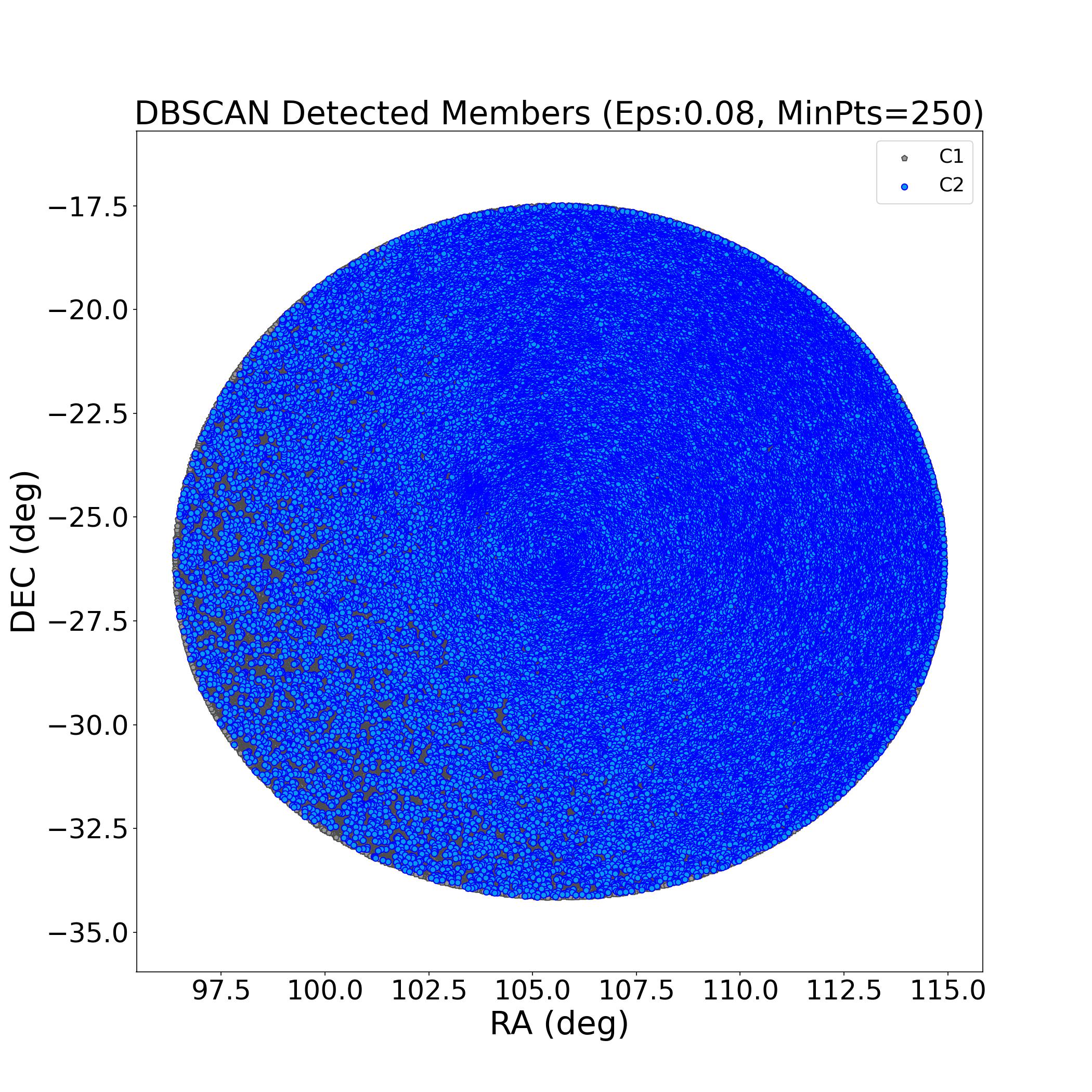}
    \end{subfigure}
    \hfill
    \begin{subfigure}{0.48\textwidth}
        \centering
        \includegraphics[width=\textwidth]{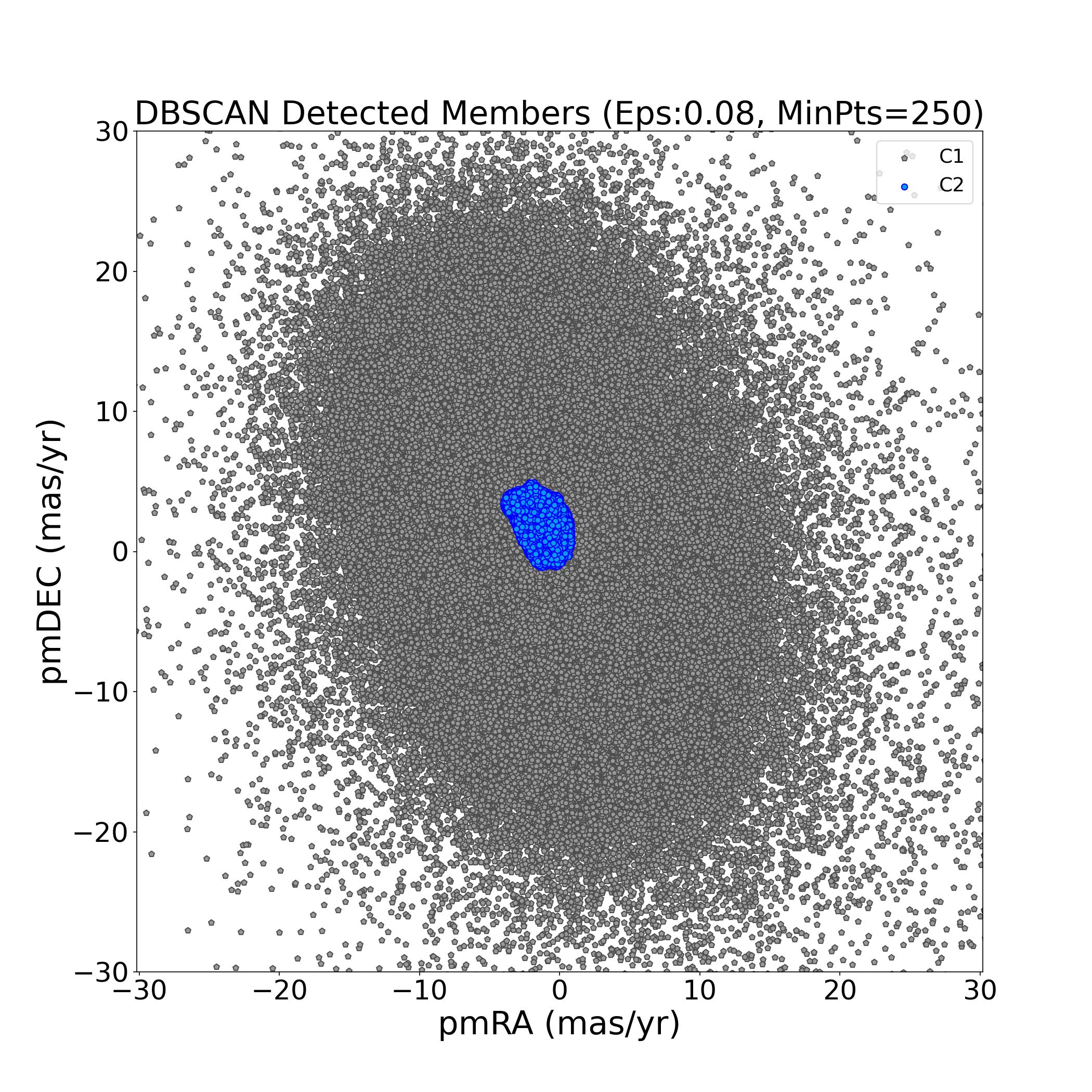}
    \end{subfigure}

    \begin{subfigure}{0.48\textwidth}
        \centering
        \includegraphics[width=\textwidth]{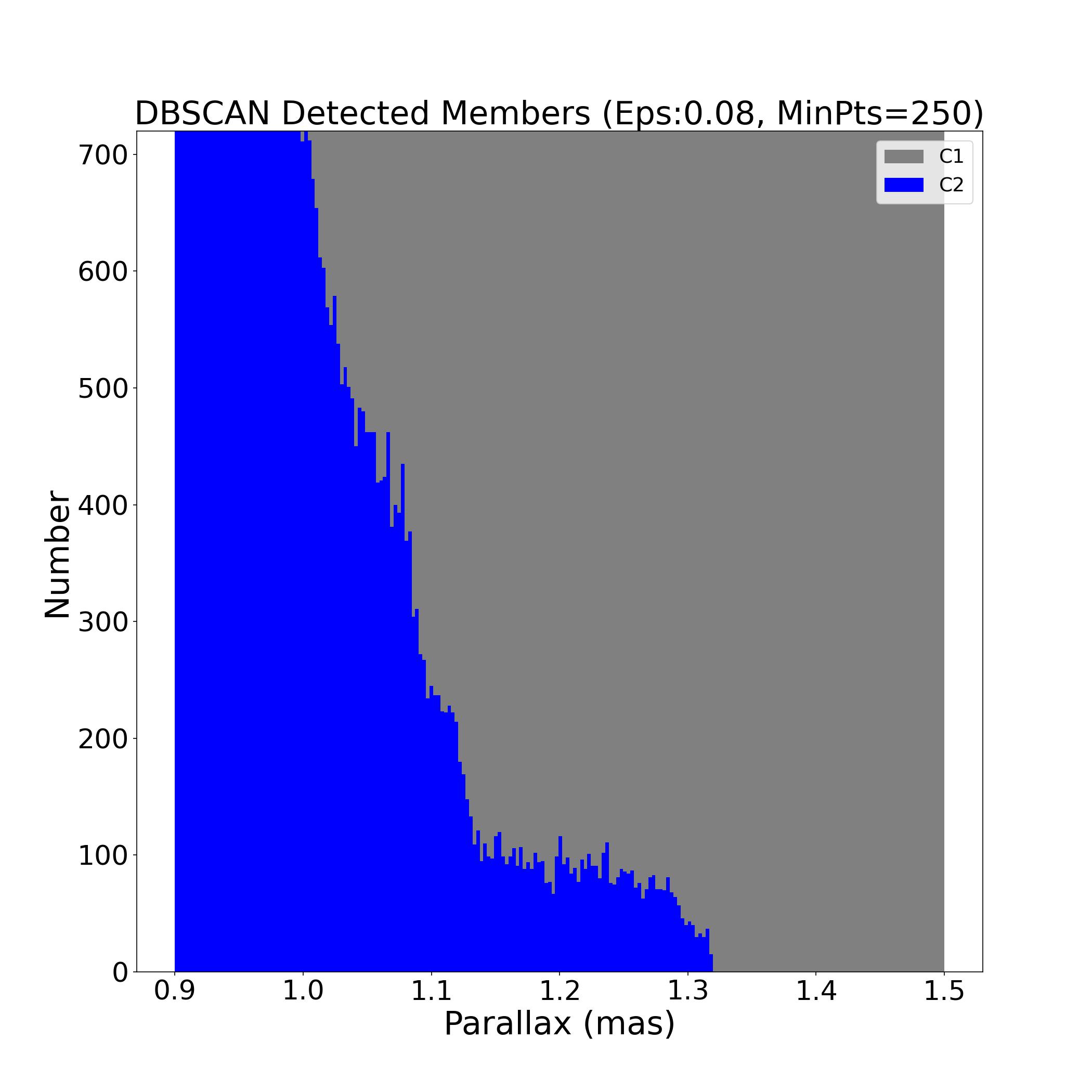}
    \end{subfigure}
    \hfill
    \begin{subfigure}{0.48\textwidth}
        \centering
        \includegraphics[width=\textwidth]{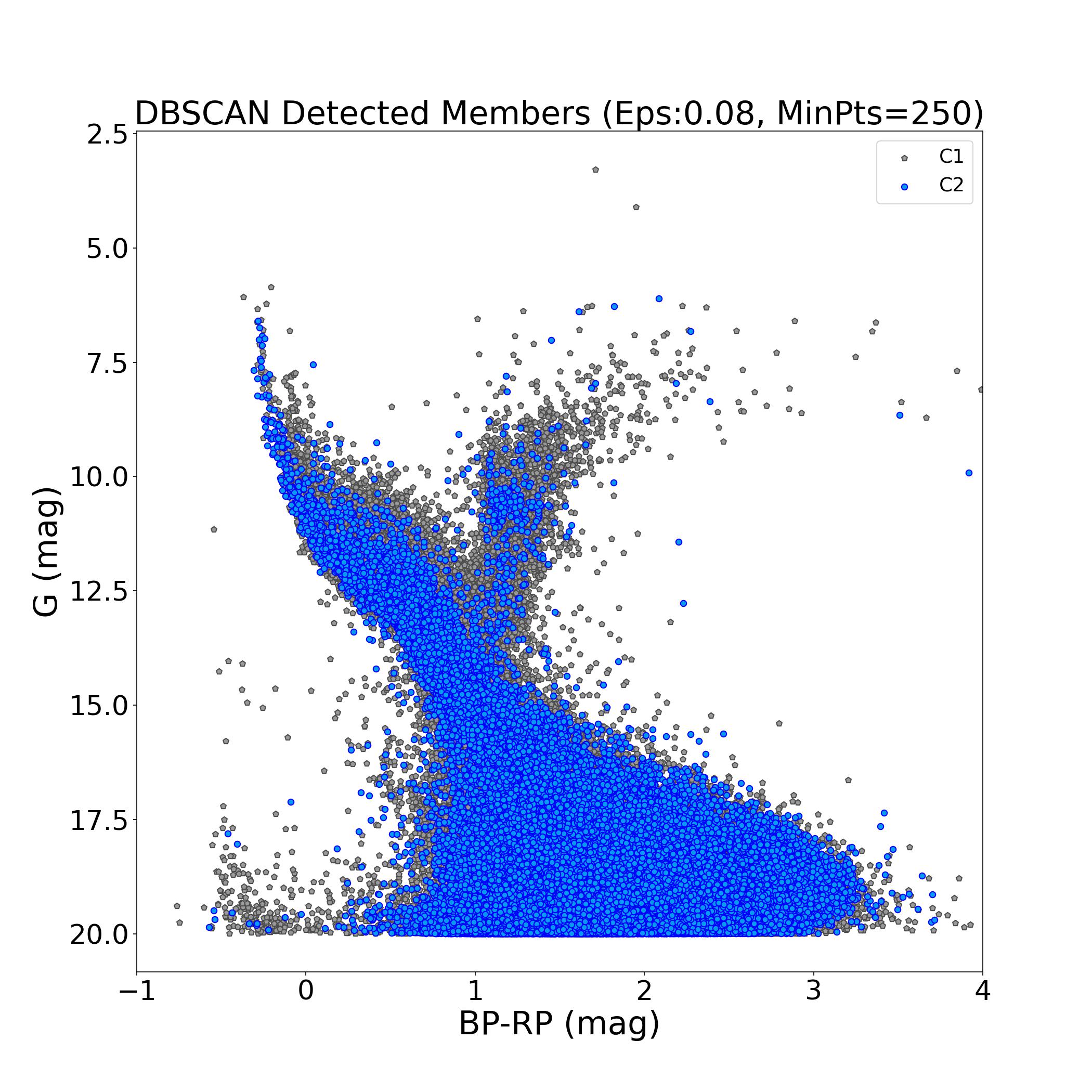}
    \end{subfigure}
    
    \caption{The DBSCAN algorithm selected candidate members based on Eps$=0.08$ and MinPts$=250$. The top-left panel illustrates the candidates' distribution in the position diagram. The top-right panel displays the distribution in the proper motion diagram, the bottom-left panel presents the parallax diagram, and the bottom-right panel shows the CMD. As observed, DBSCAN distinguishes two clusters, neither of which exhibit features typically indicative of star clusters, such as dense regions in proper motion or approximated color-magnitude diagram distribution.}
    \label{db_cluster1.fig}
\end{figure*}
\begin{figure*}
    \centering
    \begin{subfigure}{0.48\textwidth}
        \centering
        \includegraphics[width=\textwidth]{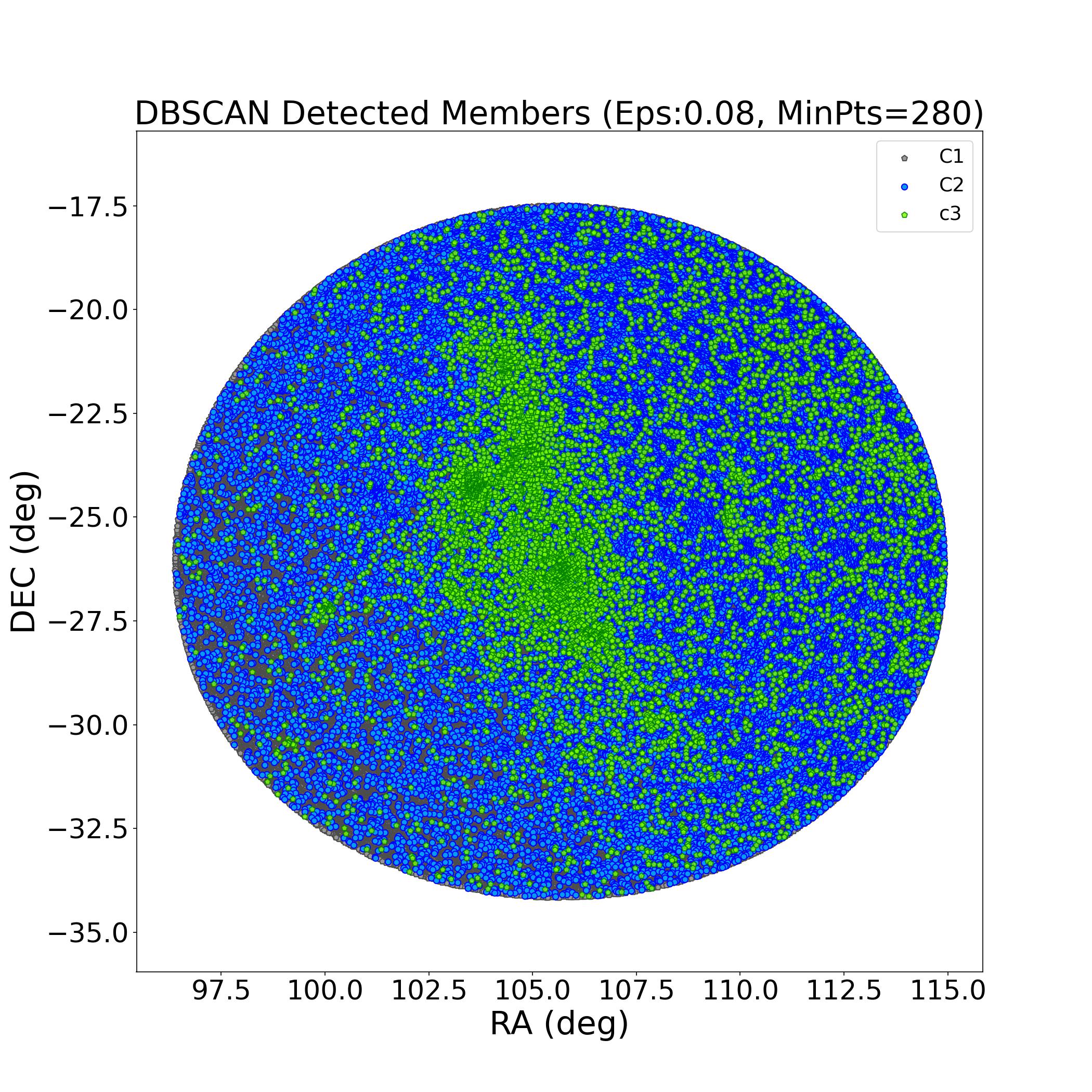}
    \end{subfigure}
    \hfill
    \begin{subfigure}{0.48\textwidth}
        \centering
        \includegraphics[width=\textwidth]{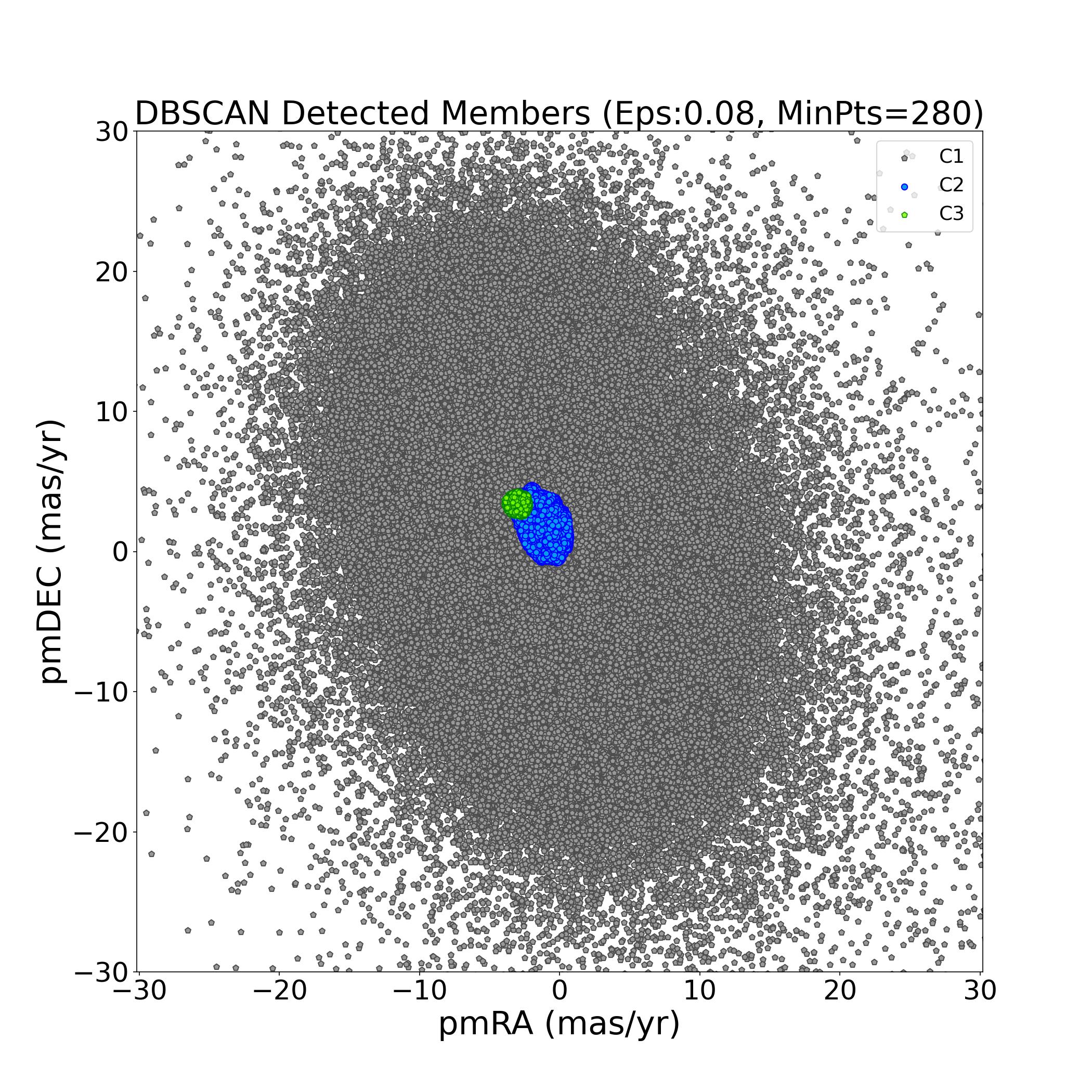}
    \end{subfigure}

    \begin{subfigure}{0.48\textwidth}
        \centering
        \includegraphics[width=\textwidth]{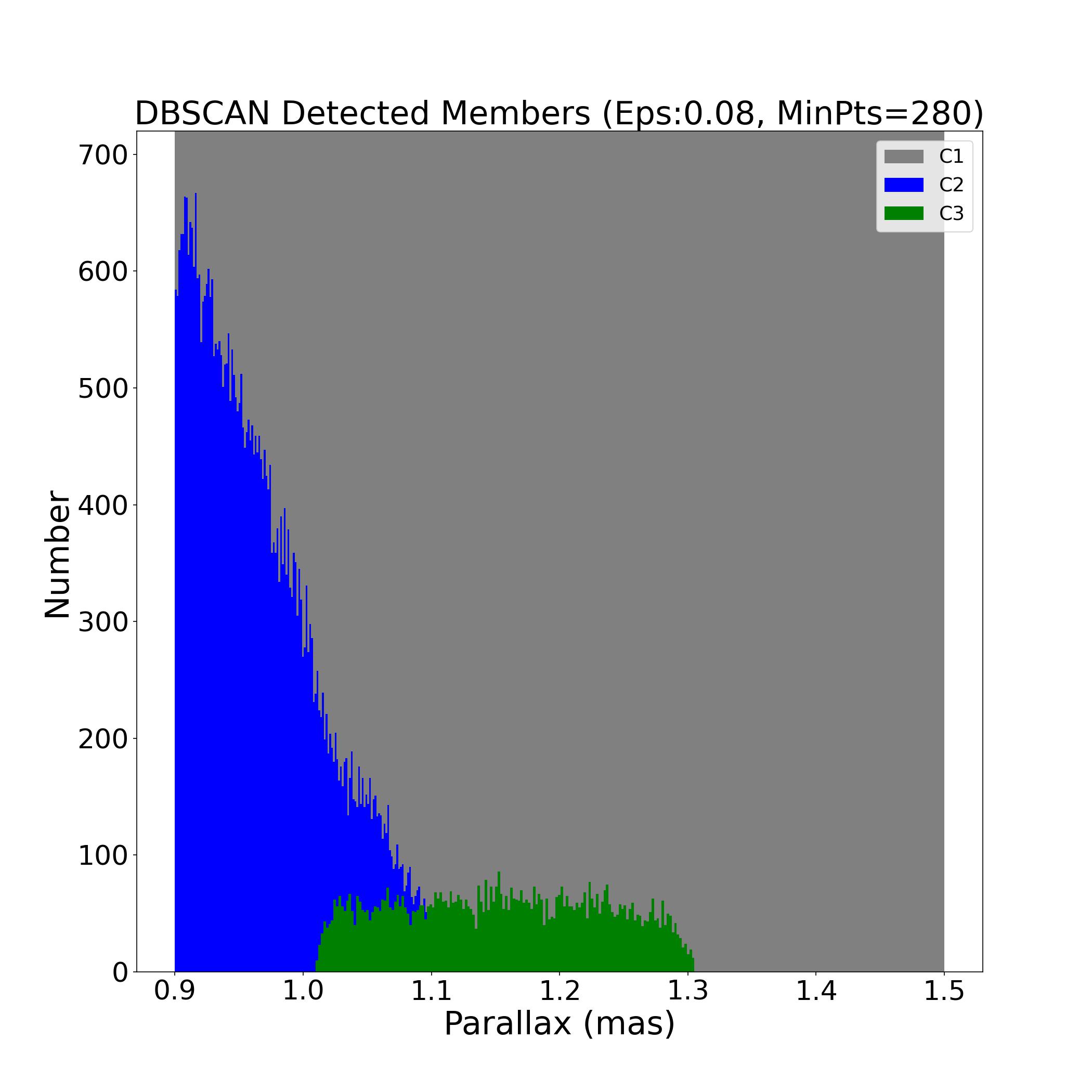}
    \end{subfigure}
    \hfill
    \begin{subfigure}{0.48\textwidth}
        \centering
        \includegraphics[width=\textwidth]{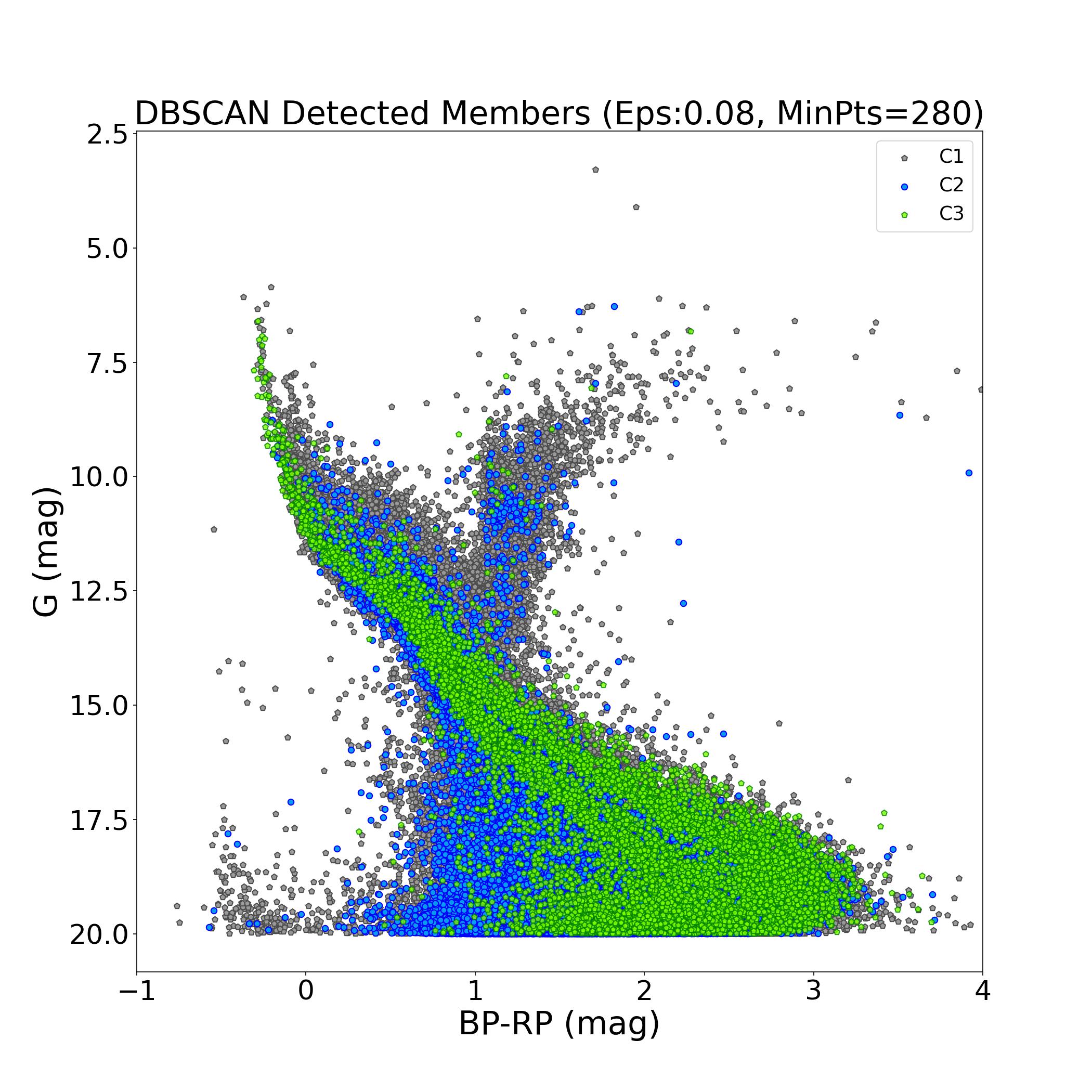}
    \end{subfigure}
    
    \caption{The DBSCAN algorithm selected candidate members based on $\mathrm{Eps}=0.08$ and $\mathrm{MinPts}=280$. With these parameters, DBSCAN identified three clusters, one of which contains our cluster members~(C\,3). The top-left panel illustrates the candidates' distribution in the position diagram, while the top-right panel presents their distribution in the proper motion diagram. The bottom-left panel displays the parallax diagram, and the bottom-right panel shows the CMD. As observed, the candidate cluster members are expected to lie within C\,3, based on the range of proper motion and parallax values. For C\,3, the CMD clearly reveals two distinct cluster distributions and indicates contamination from field stars.}
    \label{db_cluster2.fig}
\end{figure*}
\begin{figure}
  \centering
  \includegraphics[width=\linewidth]{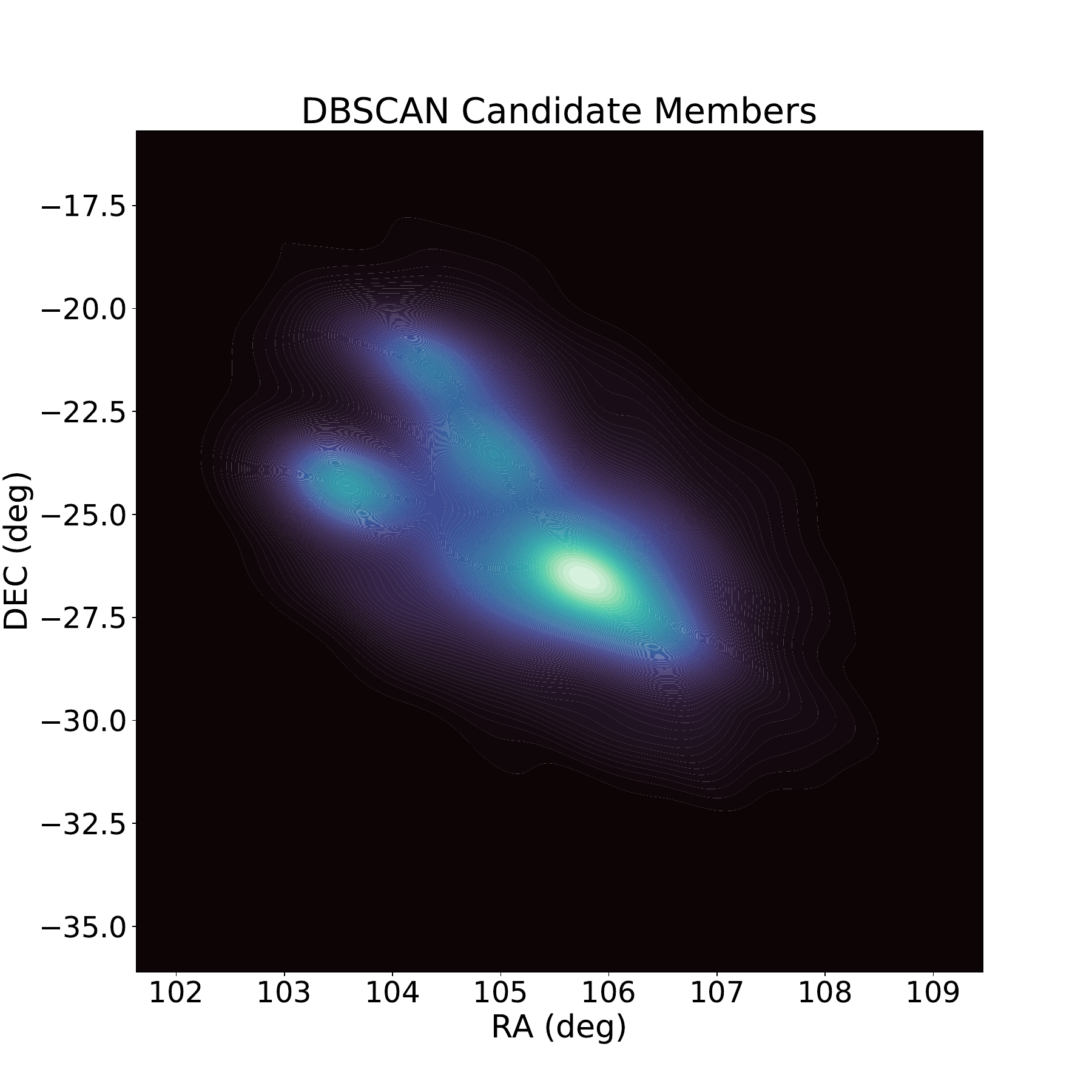}
  \caption{KDE for DBSCAN selected candidate members. This figure reveals three dense regions in the coordinates of ASCC\,32, OC\,0395 and HSC\,18.}\label{dbkde.pdf}
\end{figure}
\begin{figure*}
\centering

\begin{subfigure}{0.4\linewidth}
    \centering
    \includegraphics[width=\linewidth]{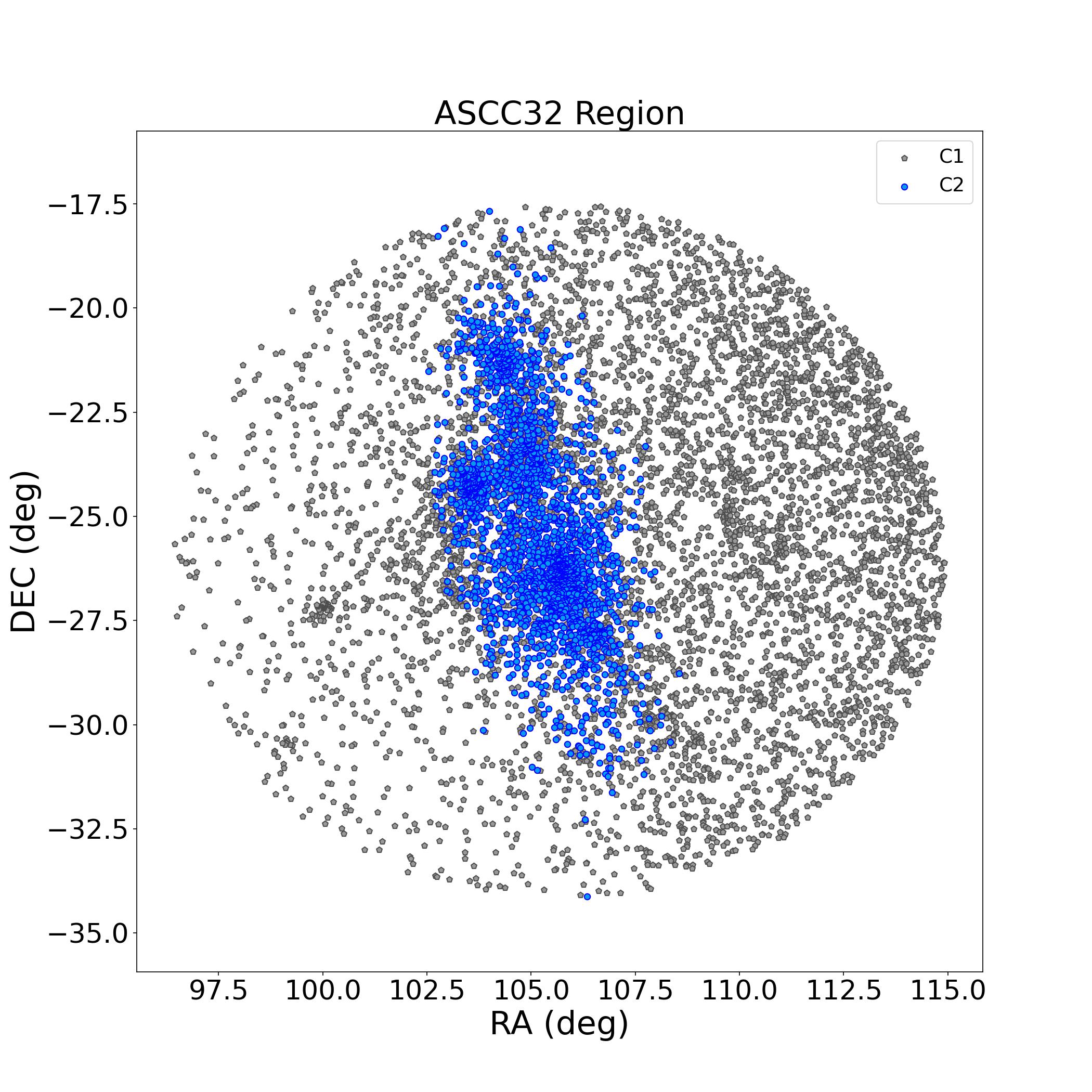}
    
\end{subfigure}
\hfill
\begin{subfigure}{0.4\linewidth}
    \centering
    \includegraphics[width=\linewidth]{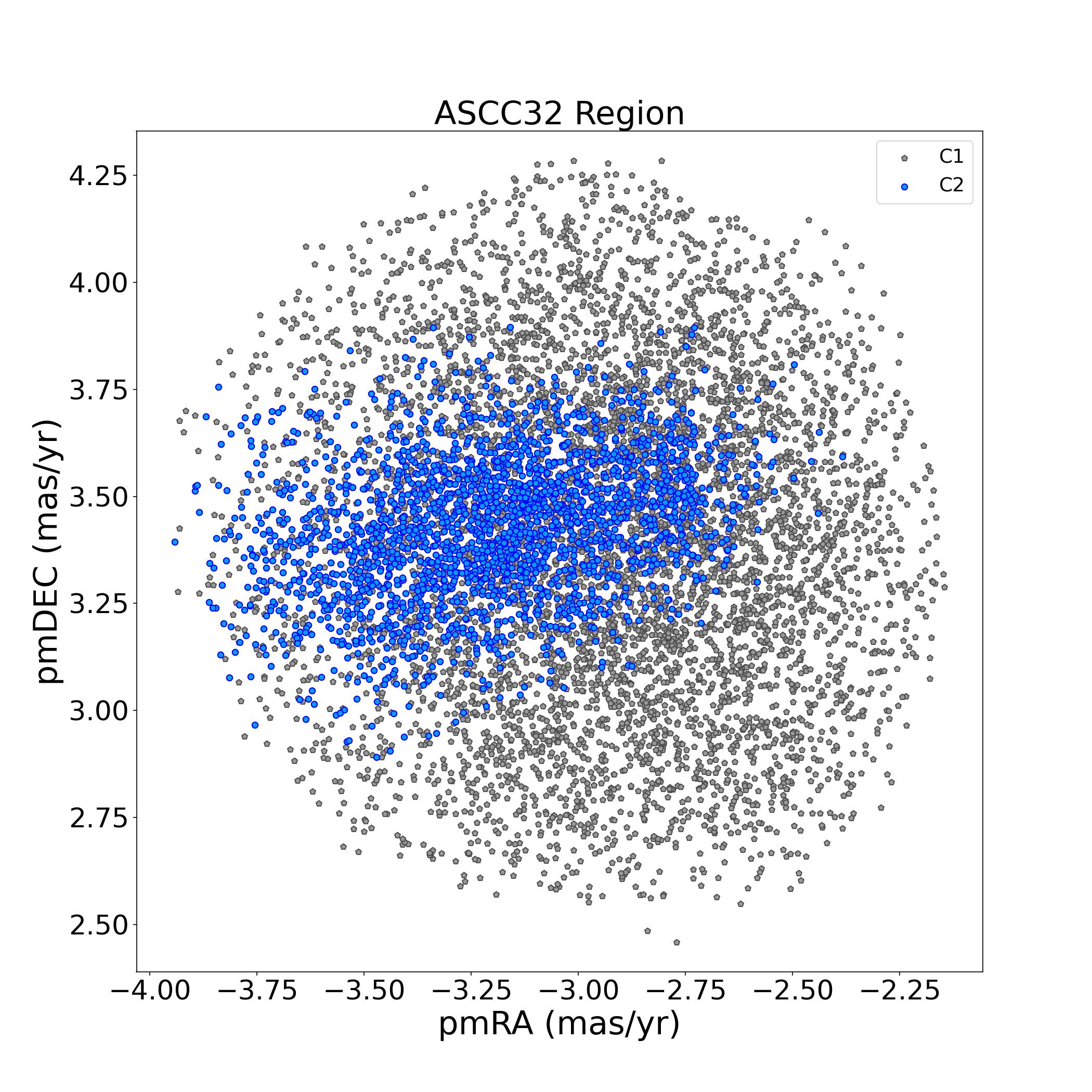}
    
\end{subfigure}

\vspace{0.5em}

\begin{subfigure}{0.4\linewidth}
    \centering
    \includegraphics[width=\linewidth]{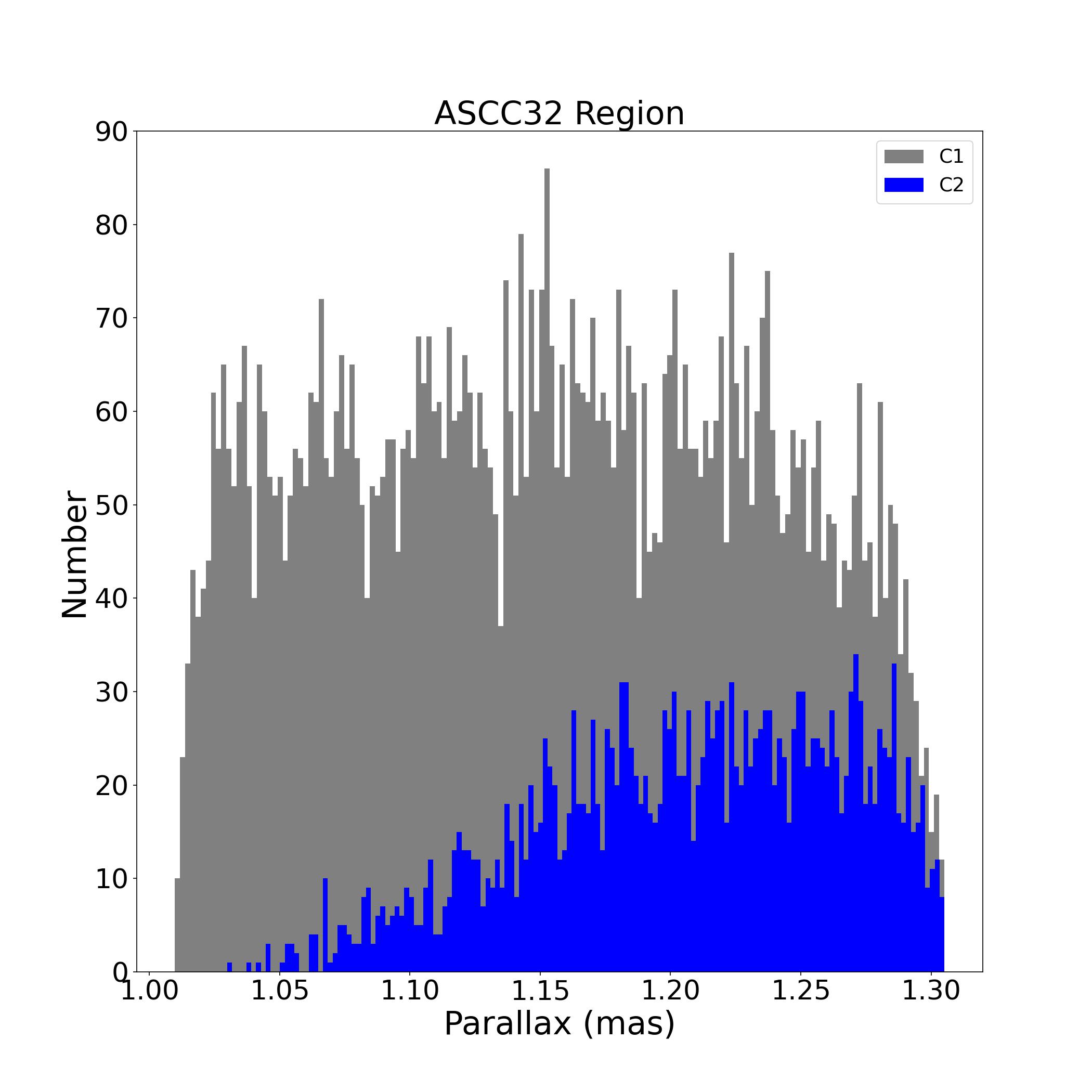}
    
\end{subfigure}
\hfill
\begin{subfigure}{0.4\linewidth}
    \centering
    \includegraphics[width=\linewidth]{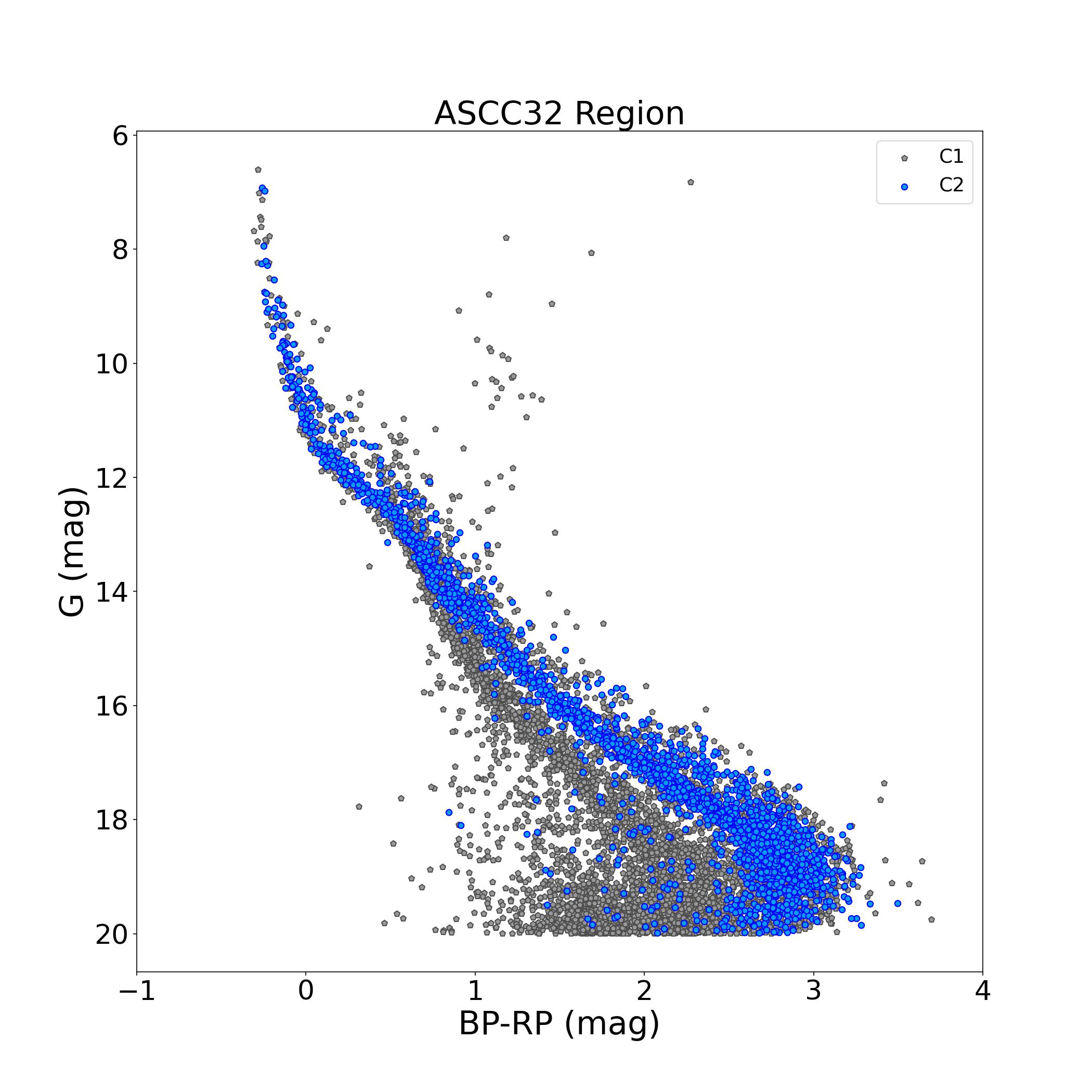}
    
\end{subfigure}

\vspace{0.5em}
\caption{Applying the GMM algorithm to cluster members detected by DG methods with a hyperparameter value of 3, one filamentary structure emerges that contains ASCC\,32, OC\,0395, and HSC\,1865. The top-left panel shows the spatial distribution, while the top-right panel illustrates the proper motion distribution. The bottom-left panel presents the parallax diagram, and the bottom-right panel displays the CMD, which demonstrates a clear main sequence without contamination.}
\label{dg_ascc}
\end{figure*}
\begin{figure}
  \centering
  \includegraphics[width=0.95\linewidth]{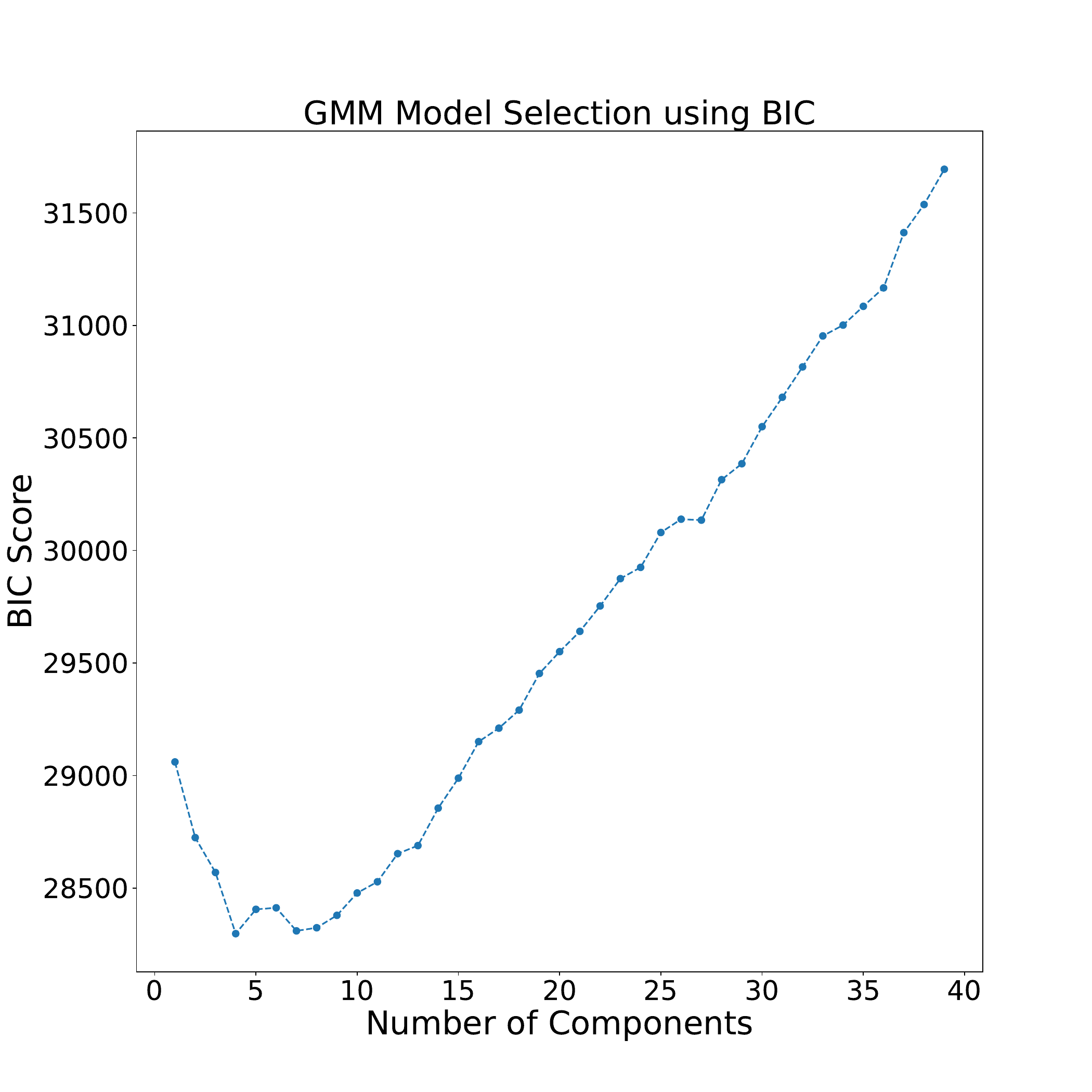}
  \caption{The BIC score determines the optimal detection parameters for GMM. At a value of 4, GMM successfully identifies reliable members for each cluster. Our analysis confirms that this value is appropriate.
\label{bic.pdf}}
\end{figure}
\begin{figure*}
\centering

\begin{subfigure}{0.4\linewidth}
    \centering
    \includegraphics[width=\linewidth]{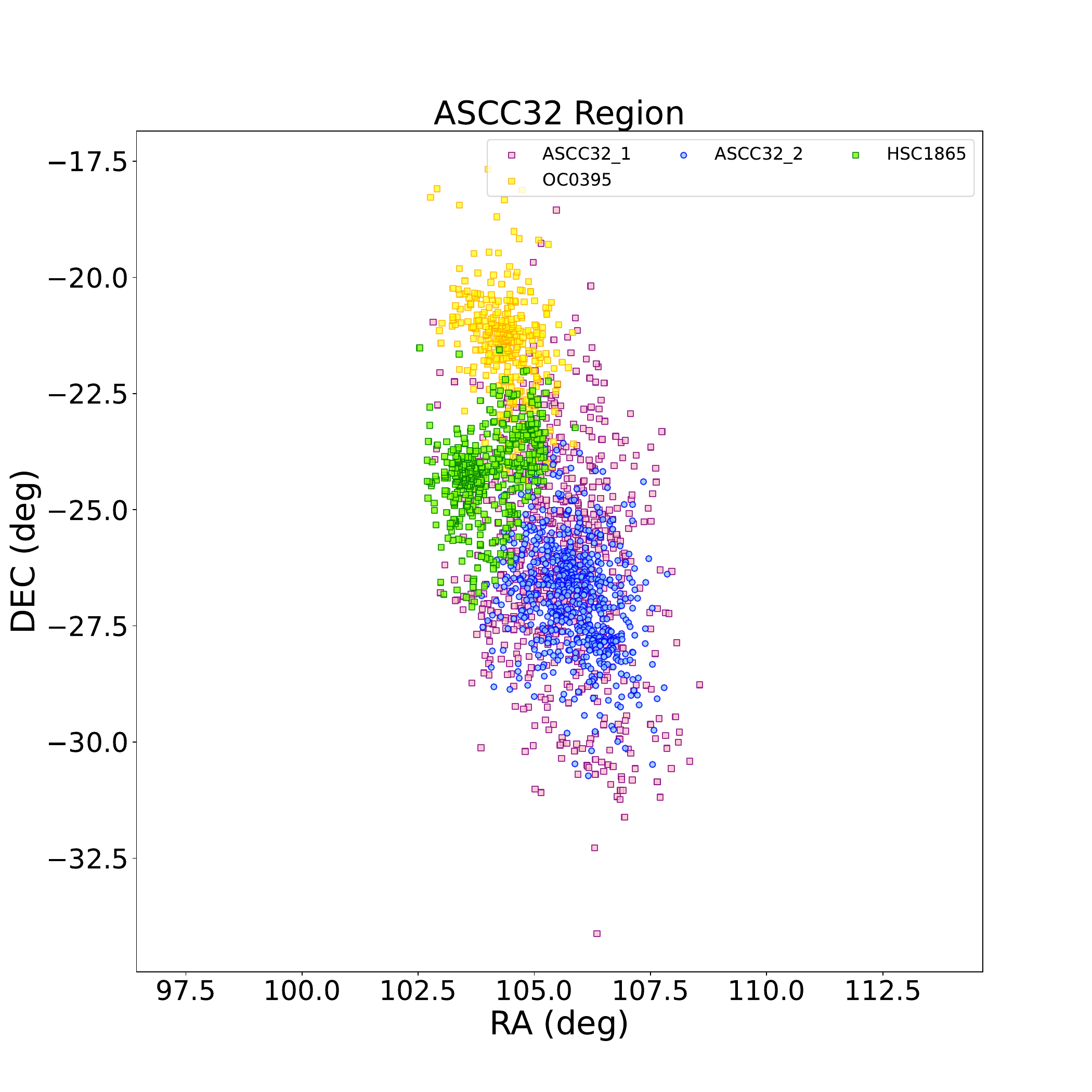}
    
\end{subfigure}
\hfill
\begin{subfigure}{0.4\linewidth}
    \centering
    \includegraphics[width=\linewidth]{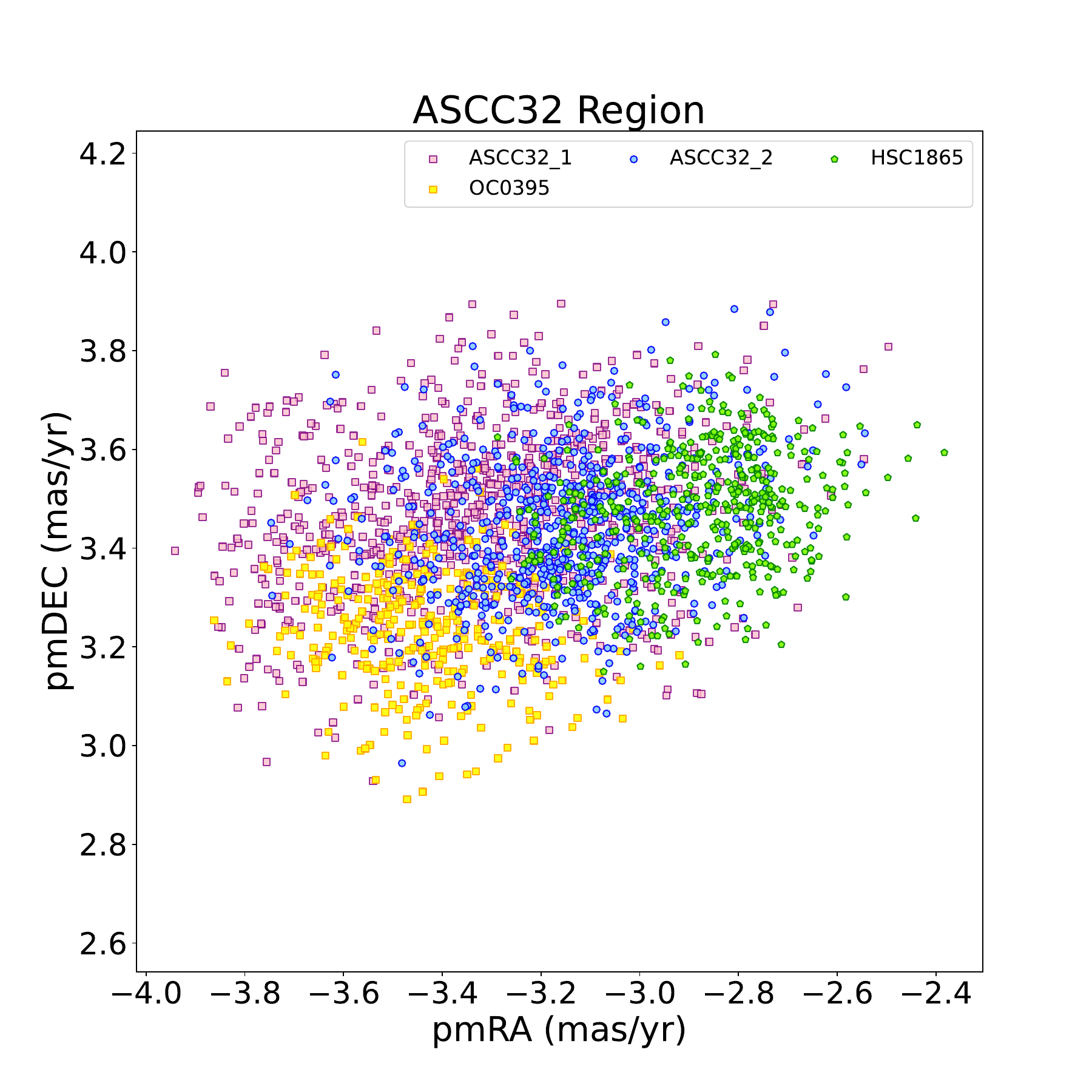}
    
\end{subfigure}

\vspace{0.5em}

\begin{subfigure}{0.4\linewidth}
    \centering
    \includegraphics[width=\linewidth]{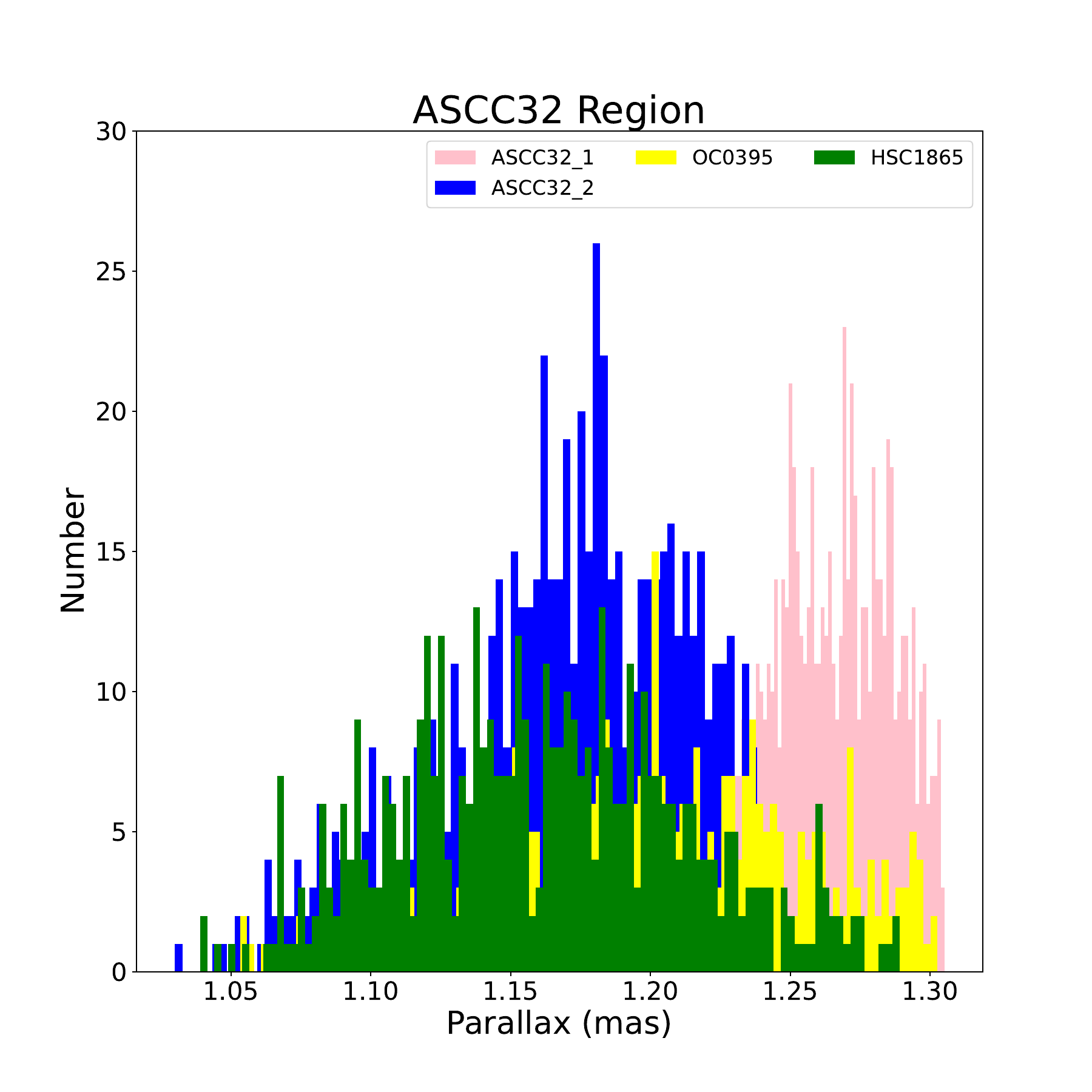}
    
\end{subfigure}
\hfill
\begin{subfigure}{0.4\linewidth}
    \centering
    \includegraphics[width=\linewidth]{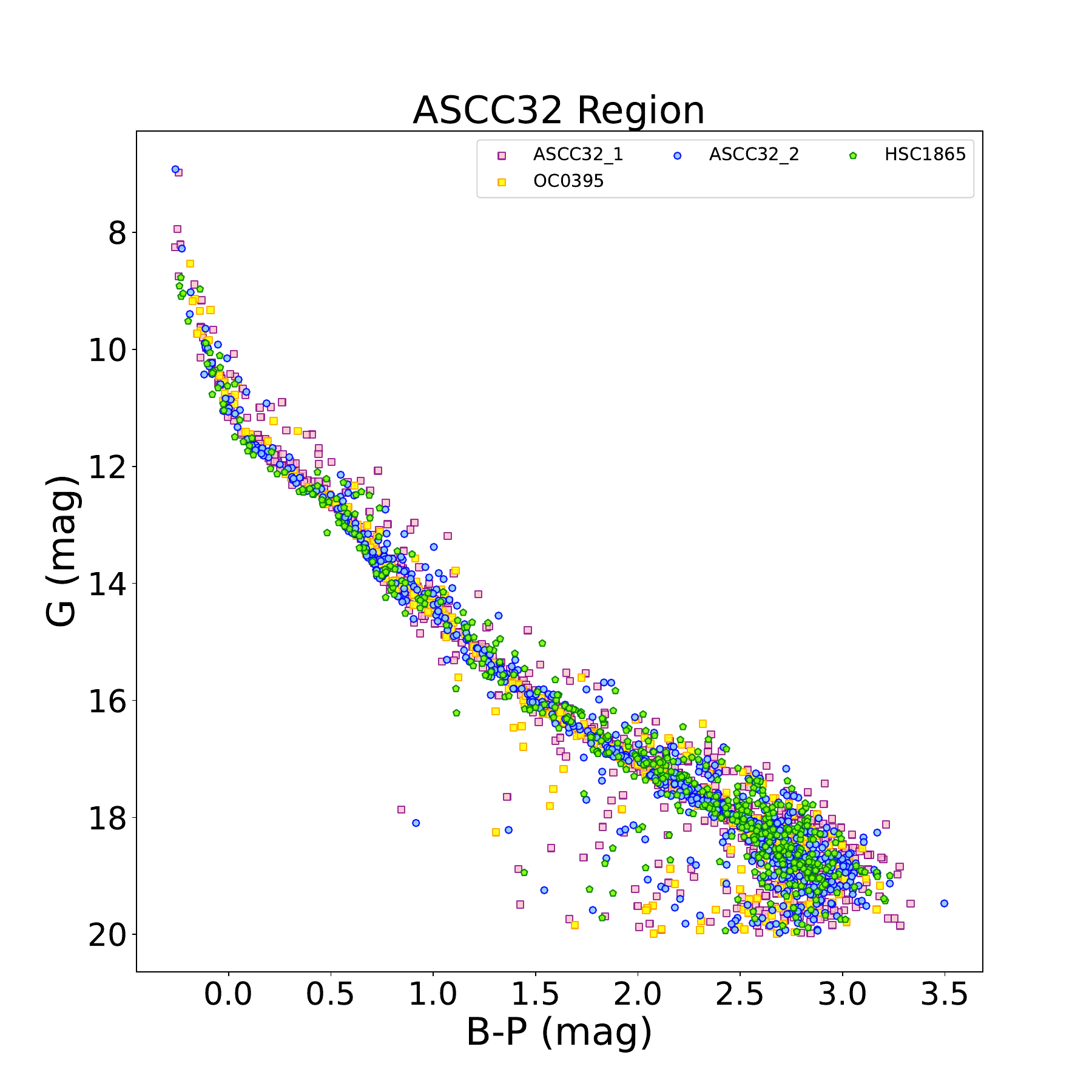}
    
\end{subfigure}

\vspace{0.5em}

\caption{Applying the GMM algorithm to filamentary structure with a hyperparameter value of 4, based on the BIC score, we identified four groups—ASCC\,32$_-$1, ASCC\,32$_-$2, OC\,0395, and HSC\,1865—each corresponding to a distinct dense region. The top-left panel shows the position distribution, while the top-right panel illustrates the proper motion distribution. The centers of each cluster exhibit slight differences. The bottom-left panel presents the parallax diagram, and the bottom-right panel displays the CMD diagram, demonstrating a clear main sequence for the four groups. These four components are spatially separated but exhibit correspondence in proper motion, parallax, and CMD, suggesting they form a Quadruple structure.}
\label{ngc_400.fig}
\end{figure*}
\begin{figure}
  \centering
  \includegraphics[width=\linewidth]{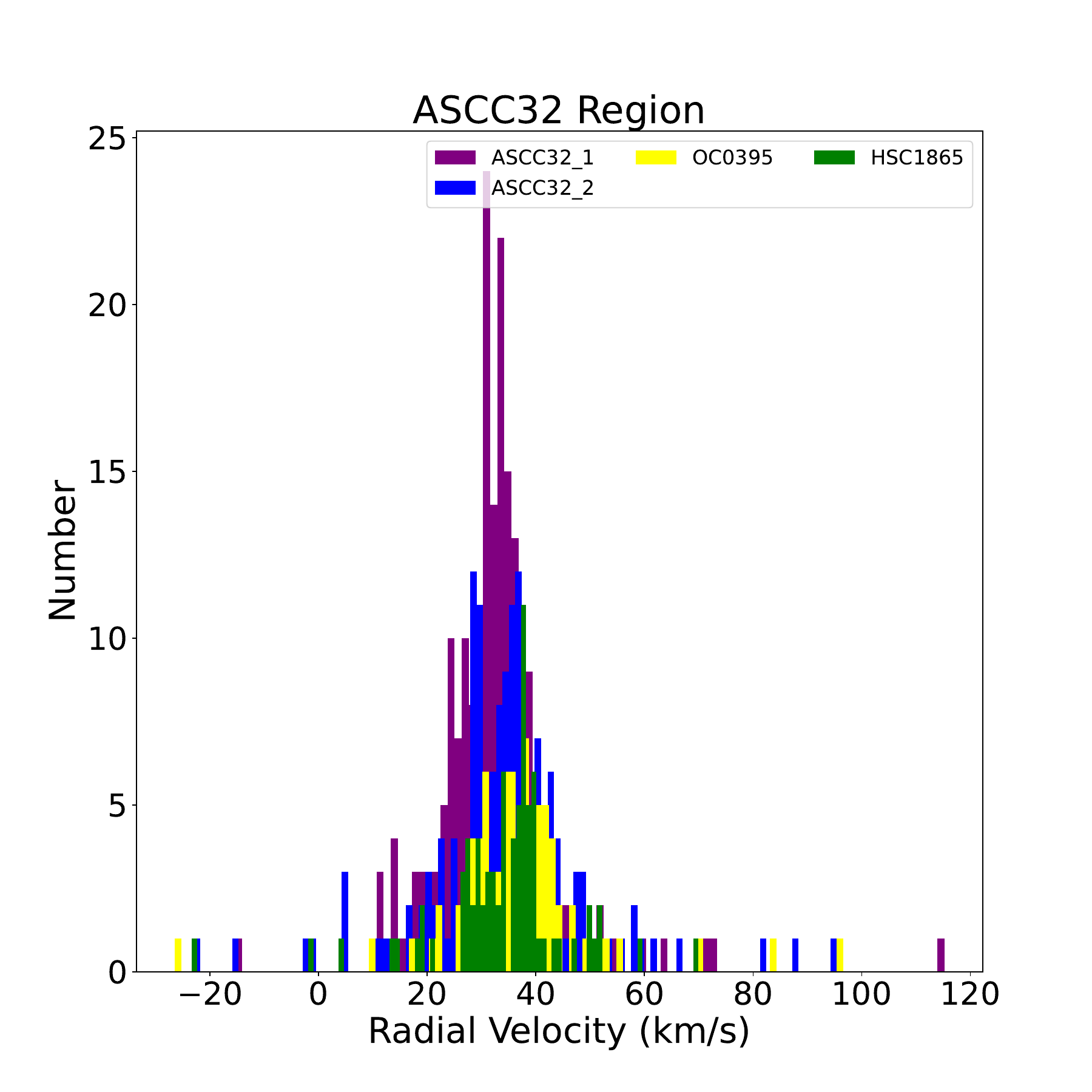}
  \caption{The radial velocity for AASCC\,32$_-$1, ASCC\,32$_-$2, OC\,0395, and HSC\,1865 is analyzed. The radial velocity distribution for all groups exhibits a similar pattern.}\label{rv.pdf}
\end{figure}

\begin{figure*}
\centering

\begin{subfigure}{0.48\linewidth}
    \centering
    \includegraphics[width=\linewidth]{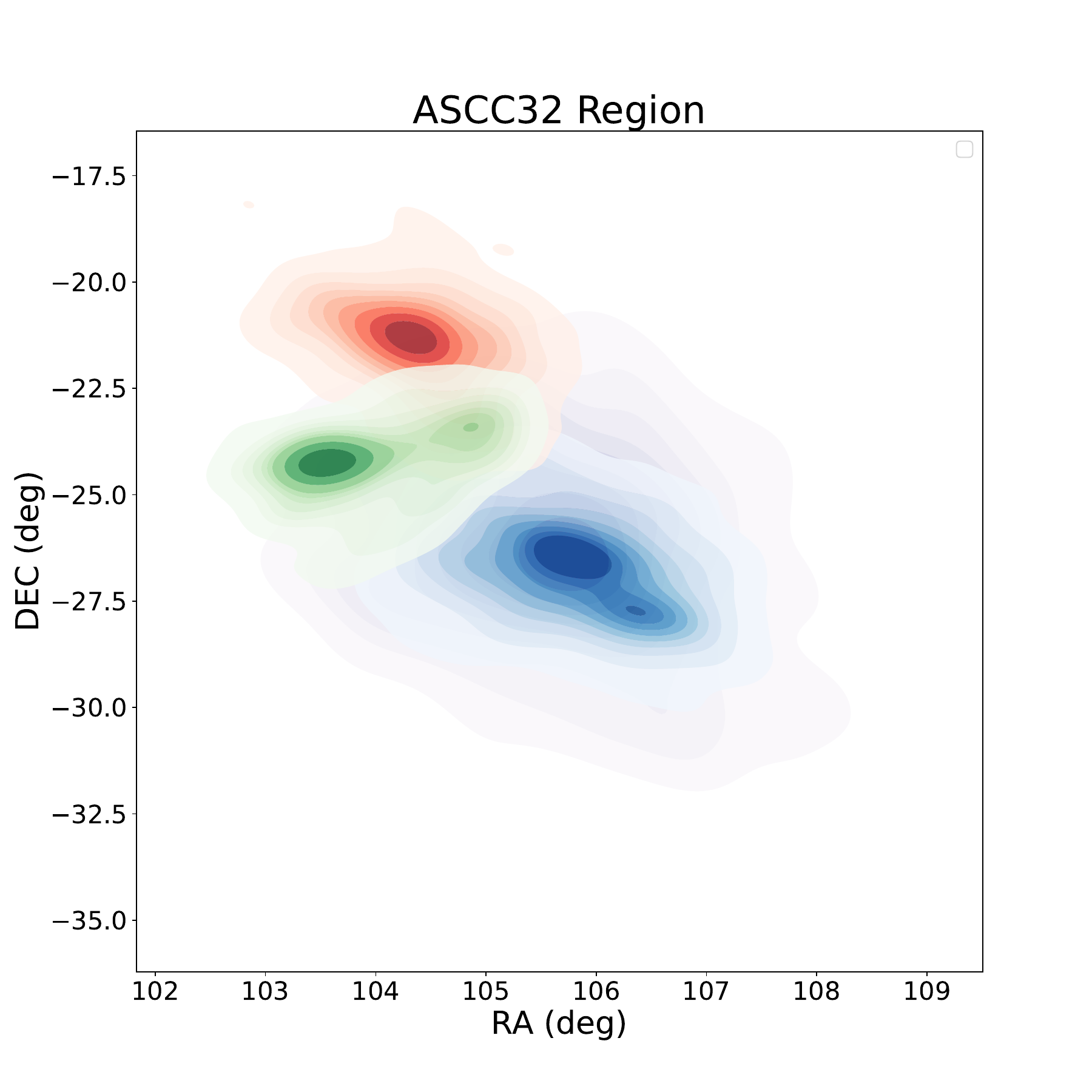}
   
\end{subfigure}
\hfill
\begin{subfigure}{0.48\linewidth}
    \centering
    \includegraphics[width=\linewidth]{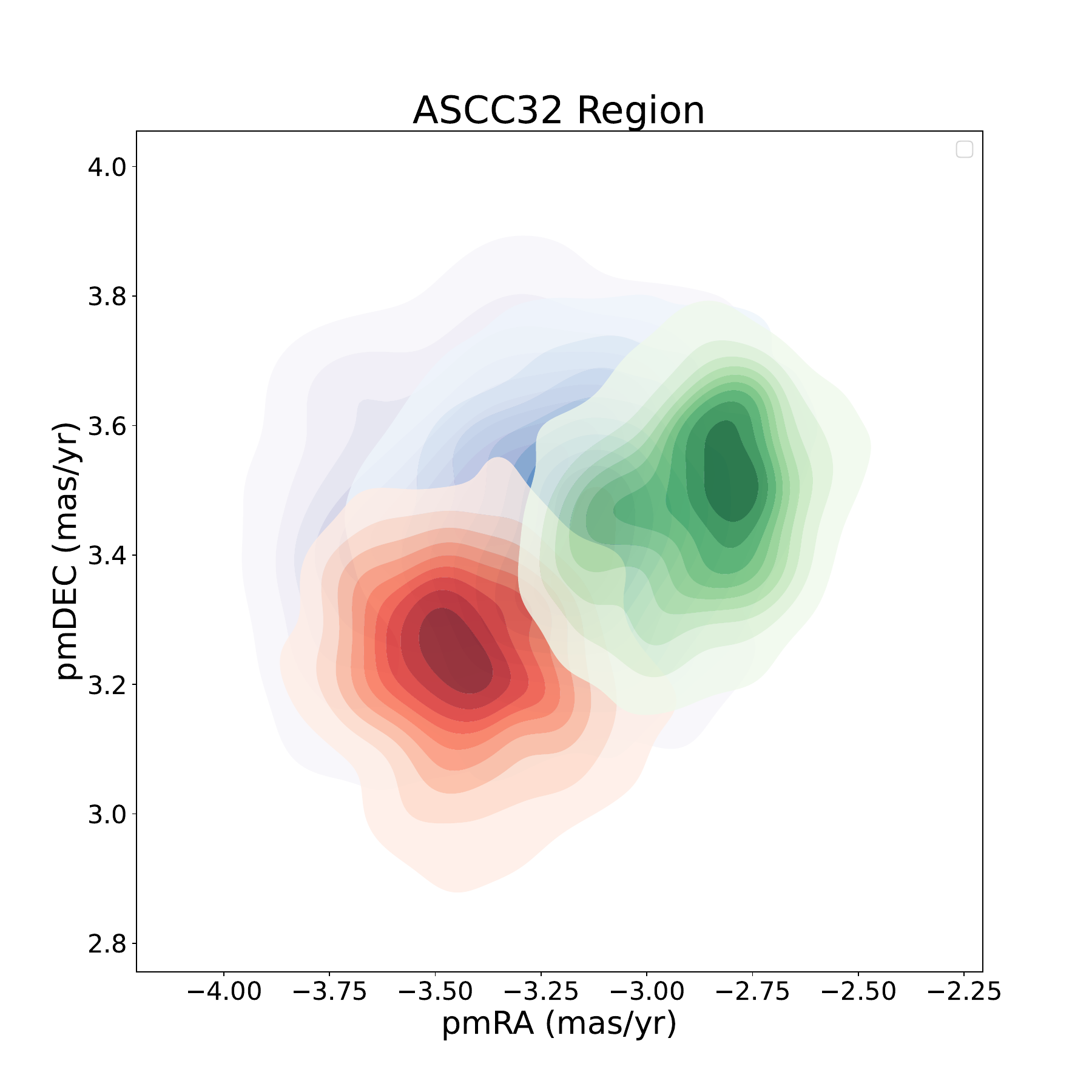}
    
\end{subfigure}
\hfill
\begin{subfigure}{0.48\linewidth}
    \centering
    \includegraphics[width=\linewidth]{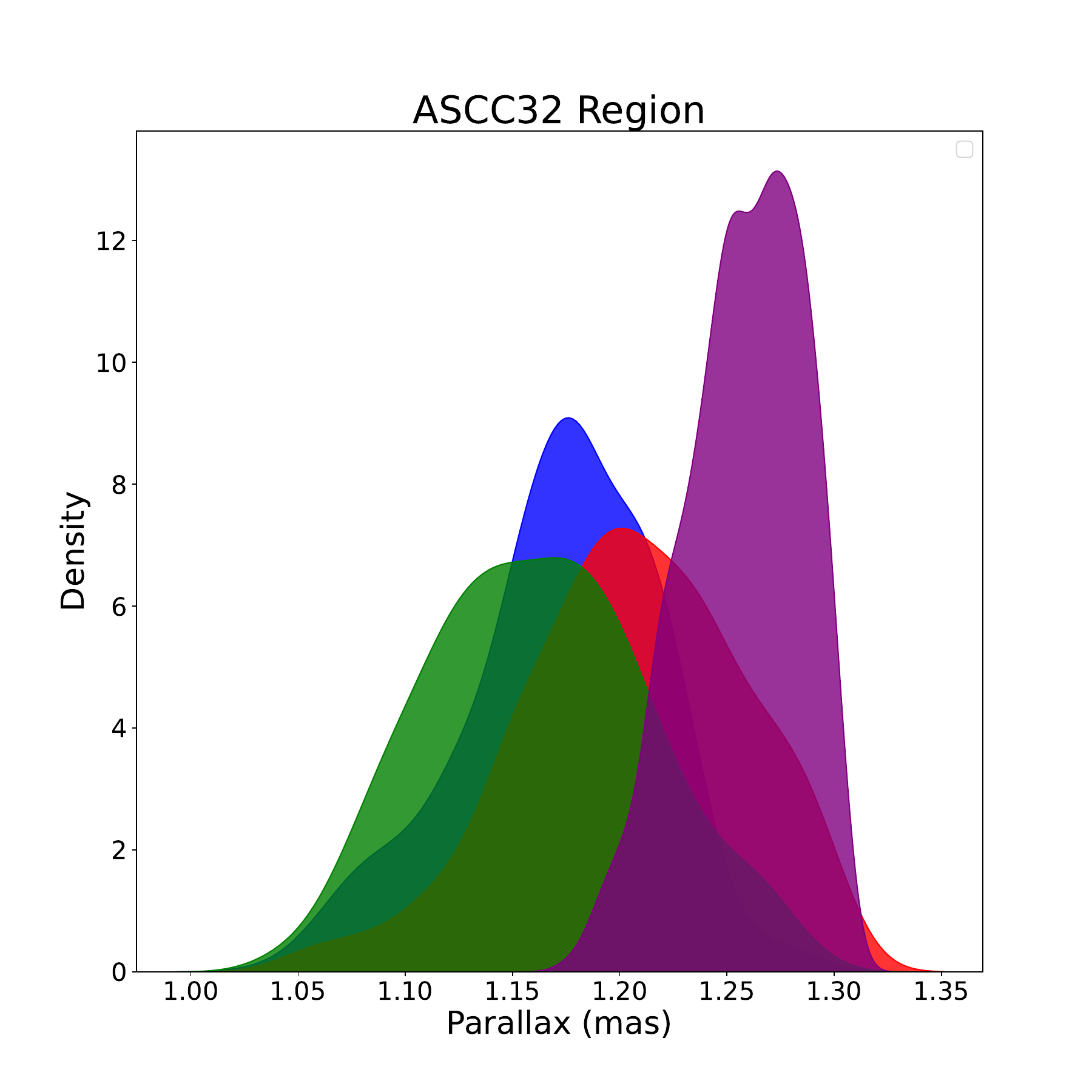}
   
\end{subfigure}

\caption{Kernel Density Estimate for ASCC\,32 region. The top-left panel represents position, while the top-right and bottom panels correspond to proper motion and parallax, respectively. Purple indicates ASCC\,32$_-$1, blue indicates ASCC\,32$_-$2, red represents OC\,0395, and green denotes HSC\,1865. As can be seen, there is a correlation between clusters based on proper motion and parallax, while their positions differ.}
\label{ngc3_kde.fig}
\end{figure*}
\begin{figure*}
\centering

\begin{subfigure}{0.48\linewidth}
    \centering
    \includegraphics[width=\linewidth]{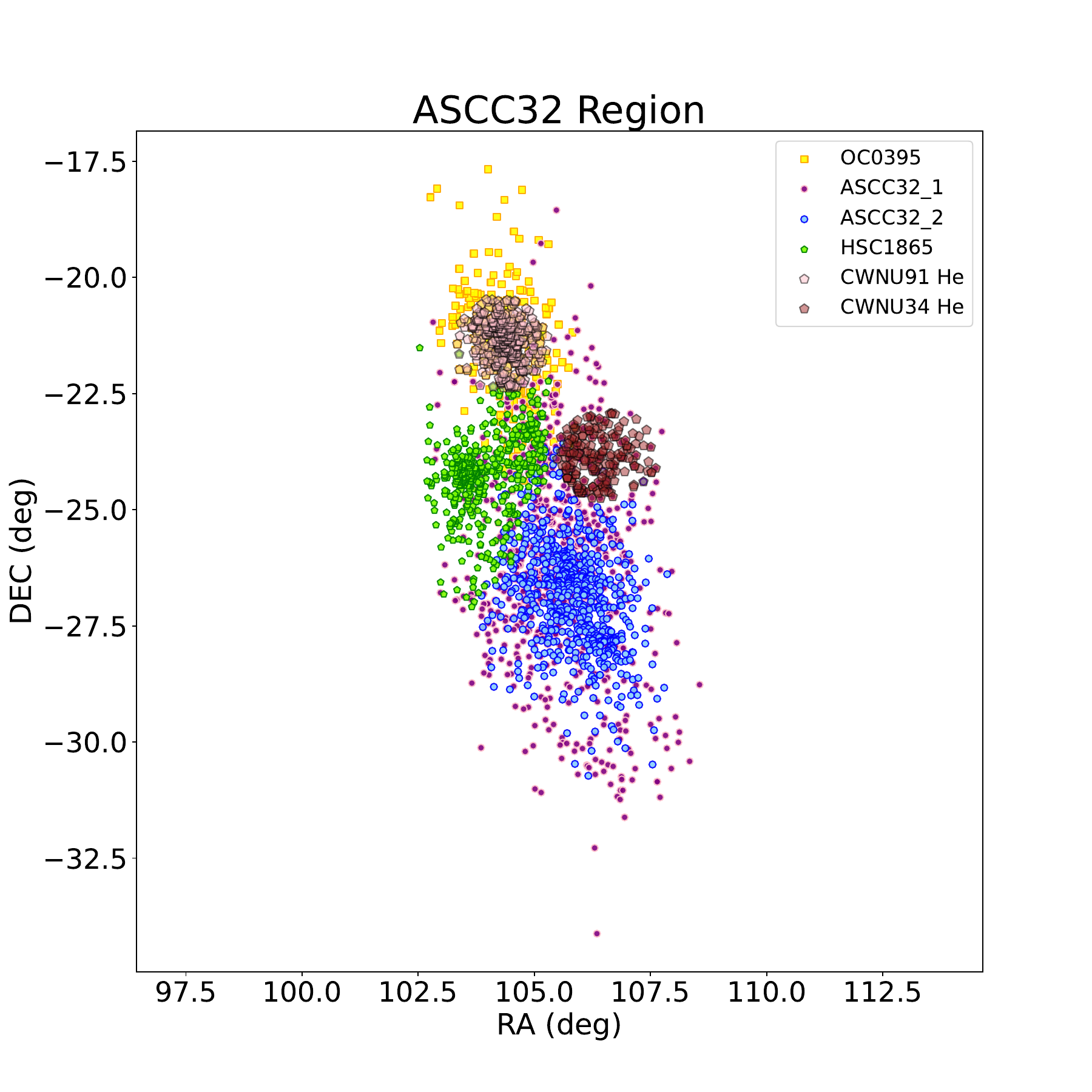}
    
\end{subfigure}
\hfill
\begin{subfigure}{0.48\linewidth}
    \centering
    \includegraphics[width=\linewidth]{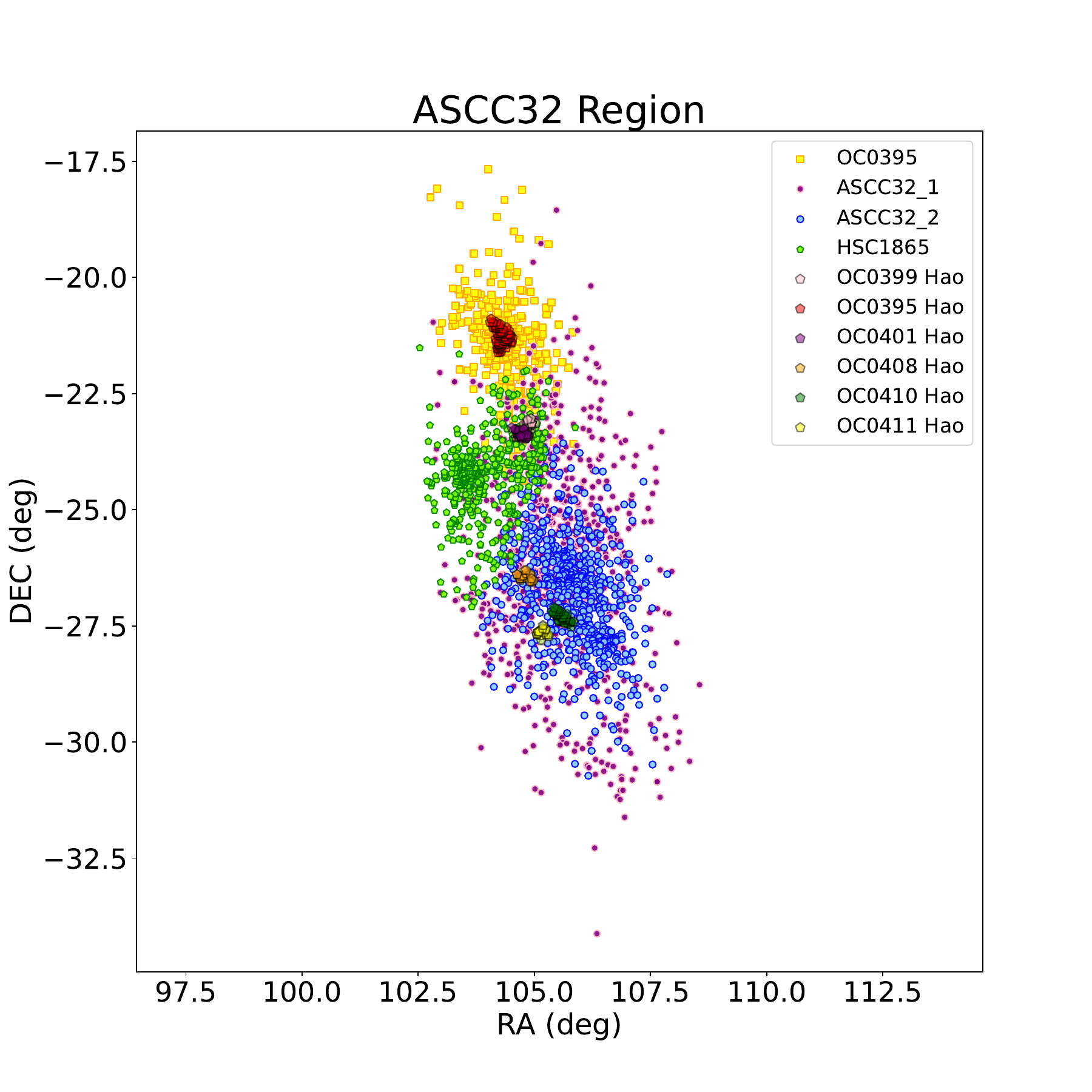}
    
\end{subfigure}

\vspace{0.5em}

\begin{subfigure}{0.48\linewidth}
    \centering
    \includegraphics[width=\linewidth]{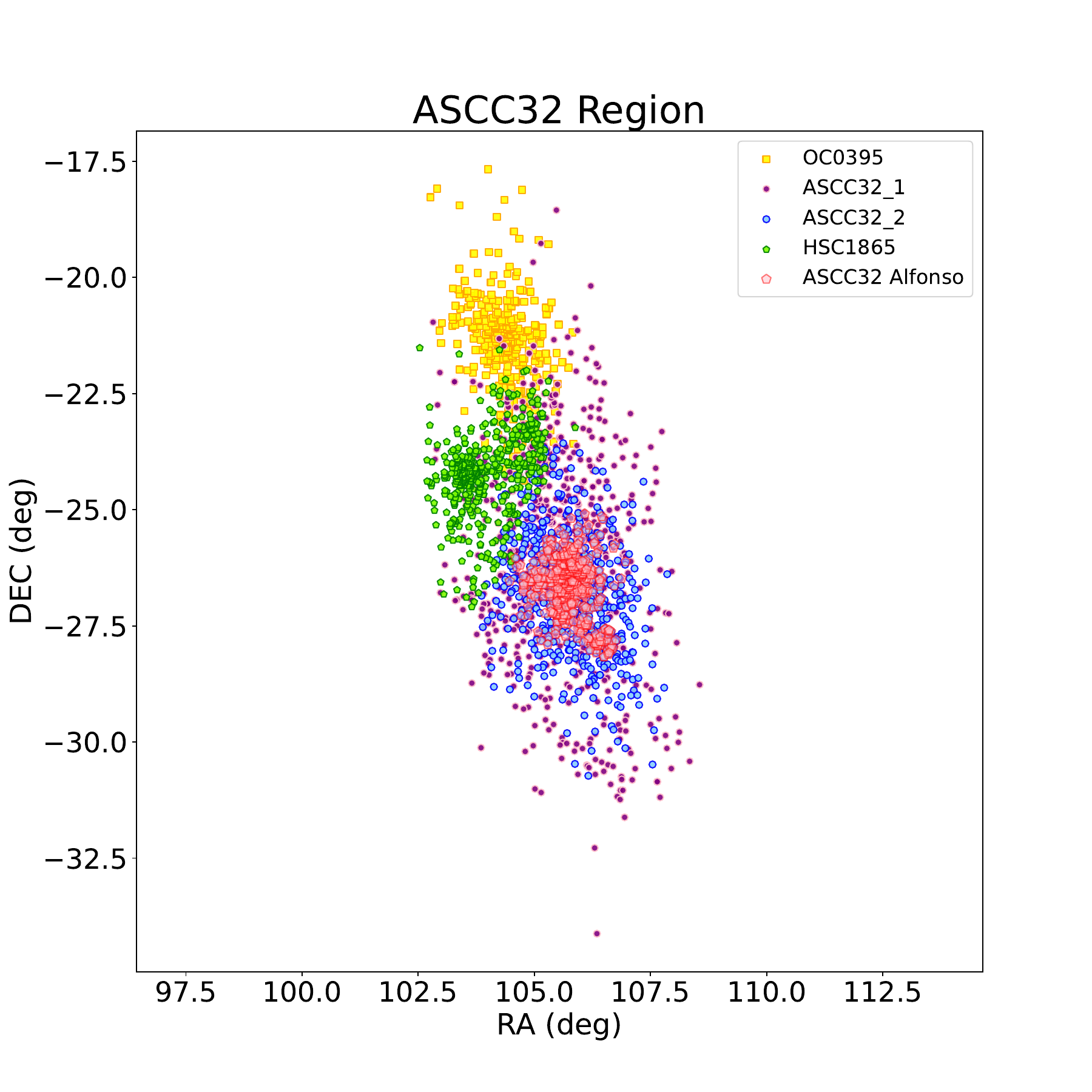}
    
\end{subfigure}
\hfill
\begin{subfigure}{0.48\linewidth}
    \centering
    \includegraphics[width=\linewidth]{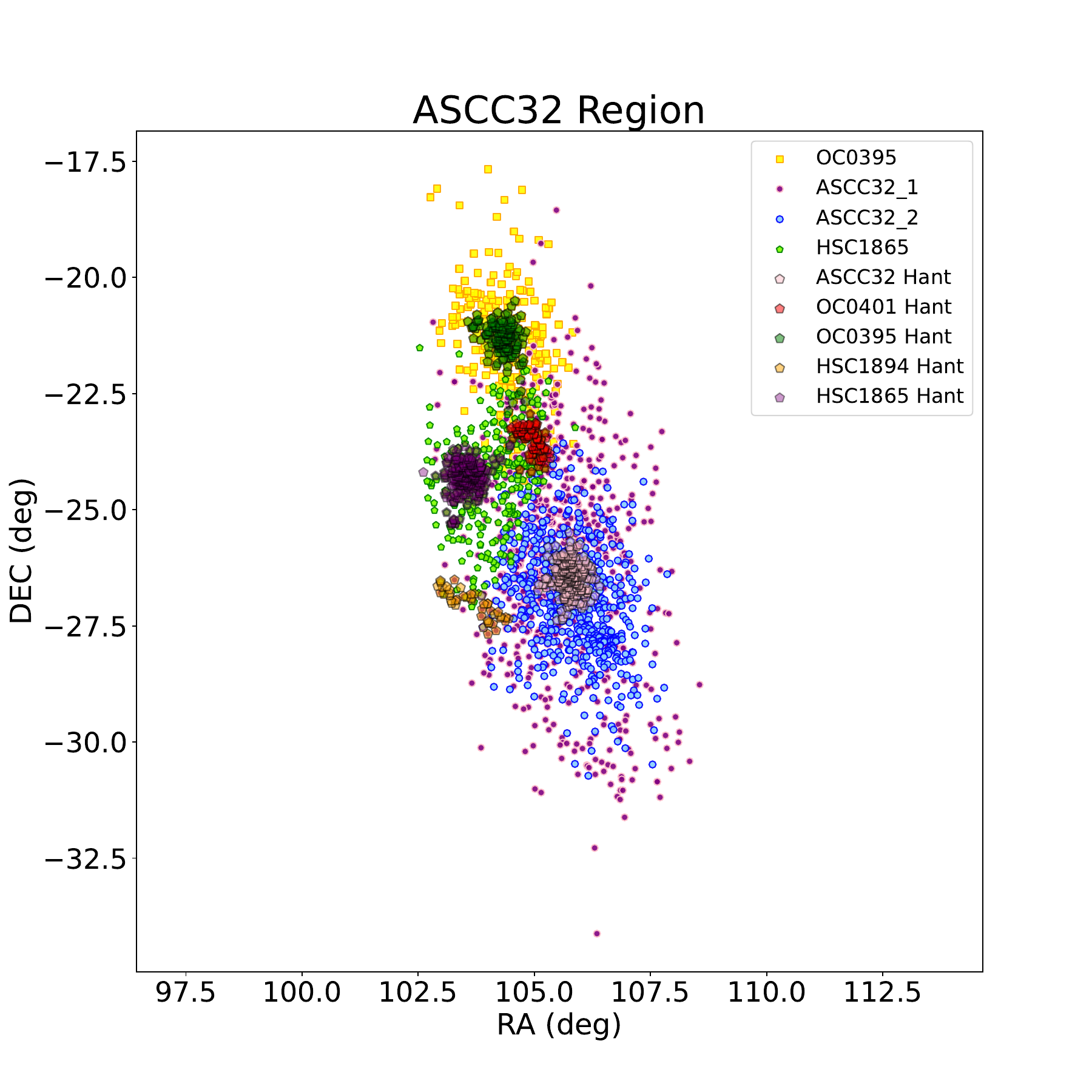}
    
\end{subfigure}

\caption{The position of each group in this work, compared against other studies The upper-left panel compares our results to He et al. (2022), while the upper-right panel compares them to Hao et al. (2022). The bottom-left panel is compared to Alfonso et al (2024), and the bottom-right panel is compared to Hunt-Reffert (2024). Our work has detected more members and provides a wide field of view from groups compared to previous studies. Some groups that were detected in earlier works are identified as part of ASCC\,32, OC\,0395, and HSC\,1865 in this study.
\label{compare.fig}}

\end{figure*}
\begin{figure*}
    \centering
        \begin{subfigure}{0.455\textwidth}
        \includegraphics[width=\textwidth]{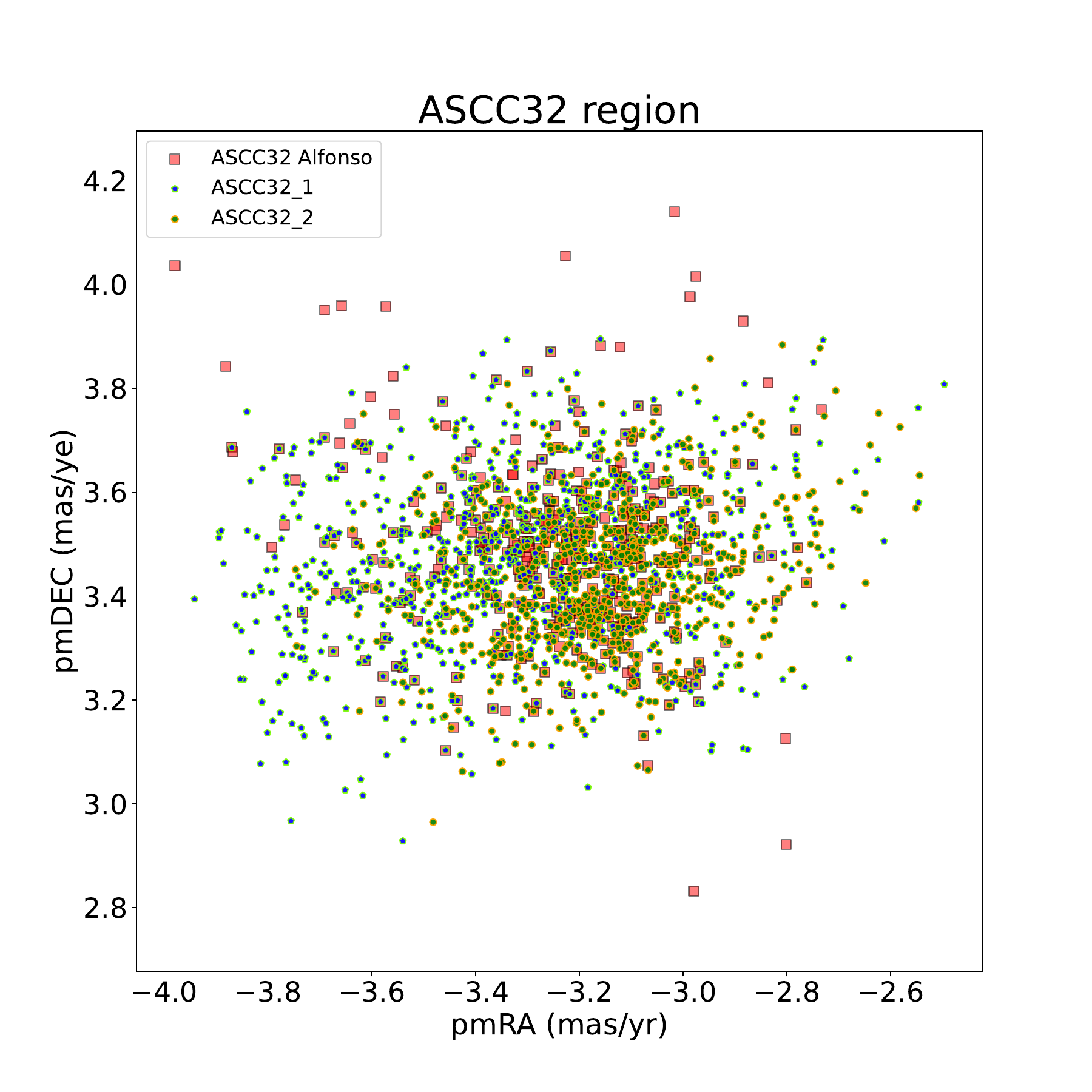}
         \end{subfigure}
        \hfill
        \begin{subfigure}{0.455\textwidth}
        \centering
        \includegraphics[width=\textwidth]{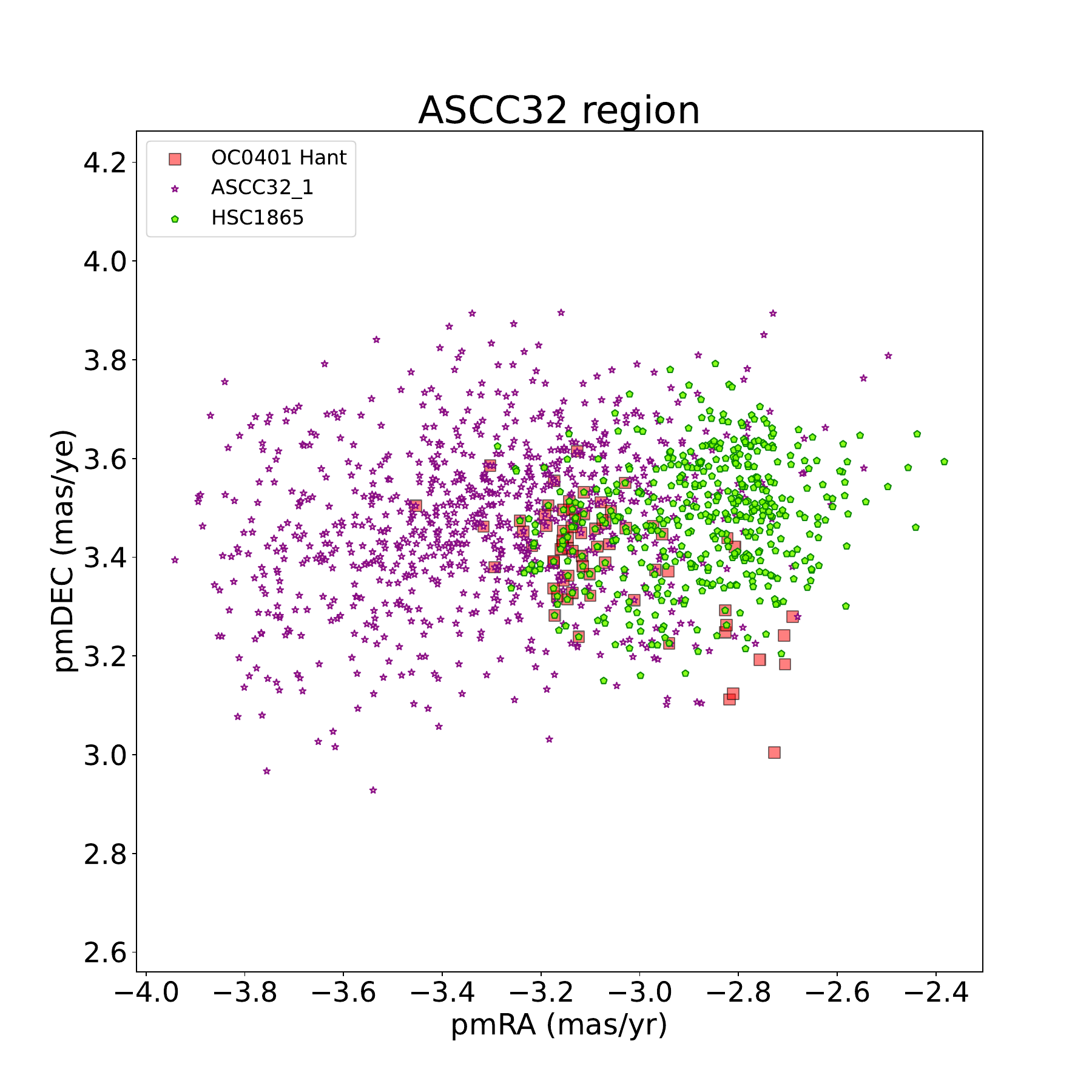}
        \end{subfigure}
    
    \caption{Comparison of proper motion distribution. The left panel compares the two distributions of ASCC\,32—ASCC\,32$_{-}$1 and ASCC\,32$_{-}$2 identified in this work—with those reported by Alfonso~(2024). As shown, the members detected in Alfonso~(2024) correspond to the two distributions of ASCC\,32 presented in this study. The right panel compares HSC\,1865 and ASCC\,32$_{-}$1 from this work with OC\,0401 from Hunt and Reffert~(2024), which was identified as part of both HSC\,1865 and ASCC\,32$_{-}$1 in our analysis.}
    \label{pmc_co.fig}
\end{figure*}
\begin{figure*}
\centering

\begin{subfigure}{0.49\linewidth}
    \centering
    \includegraphics[width=\linewidth]{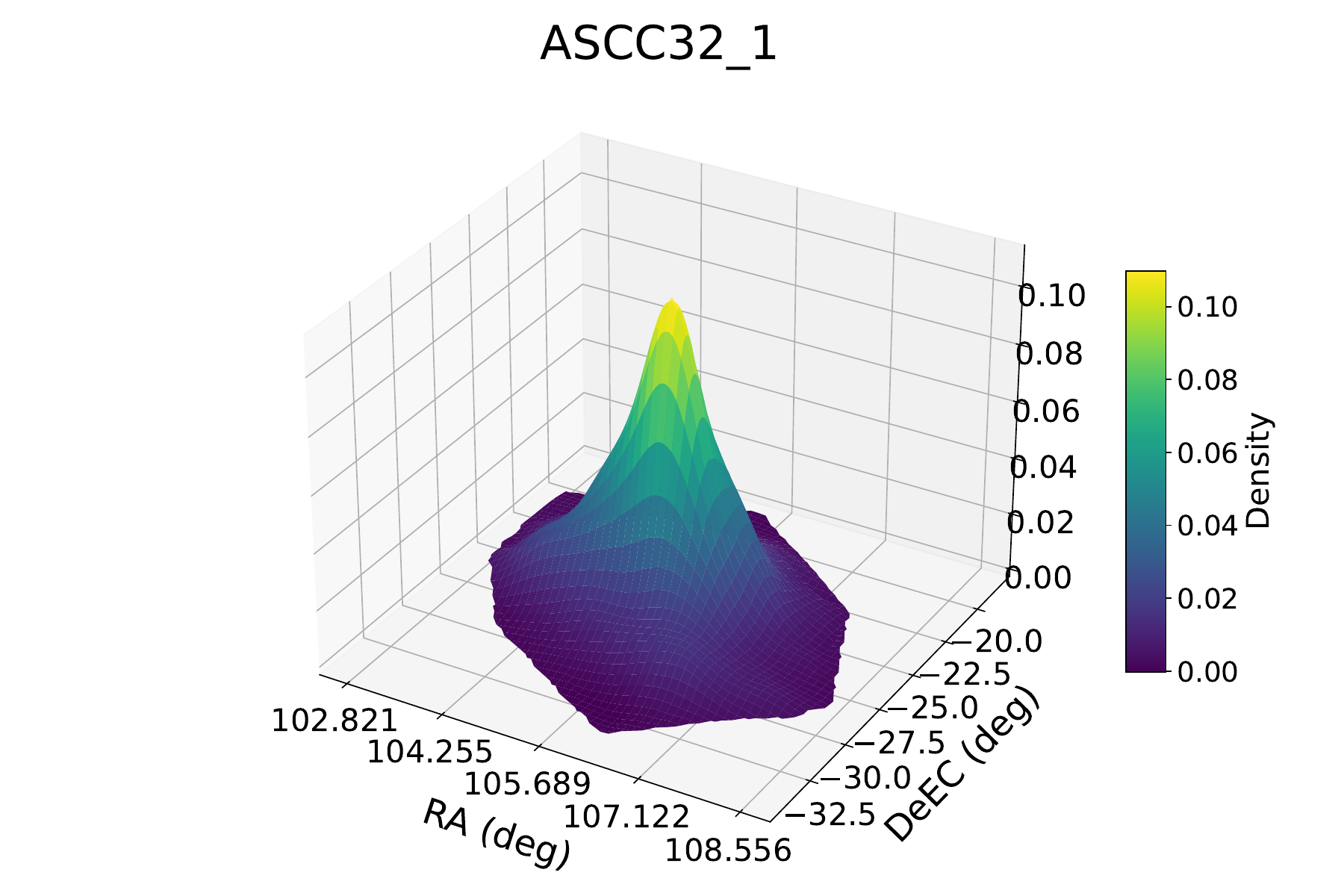}
    
\end{subfigure}
\hfill
\begin{subfigure}{0.49\linewidth}
    \centering
    \includegraphics[width=\linewidth]{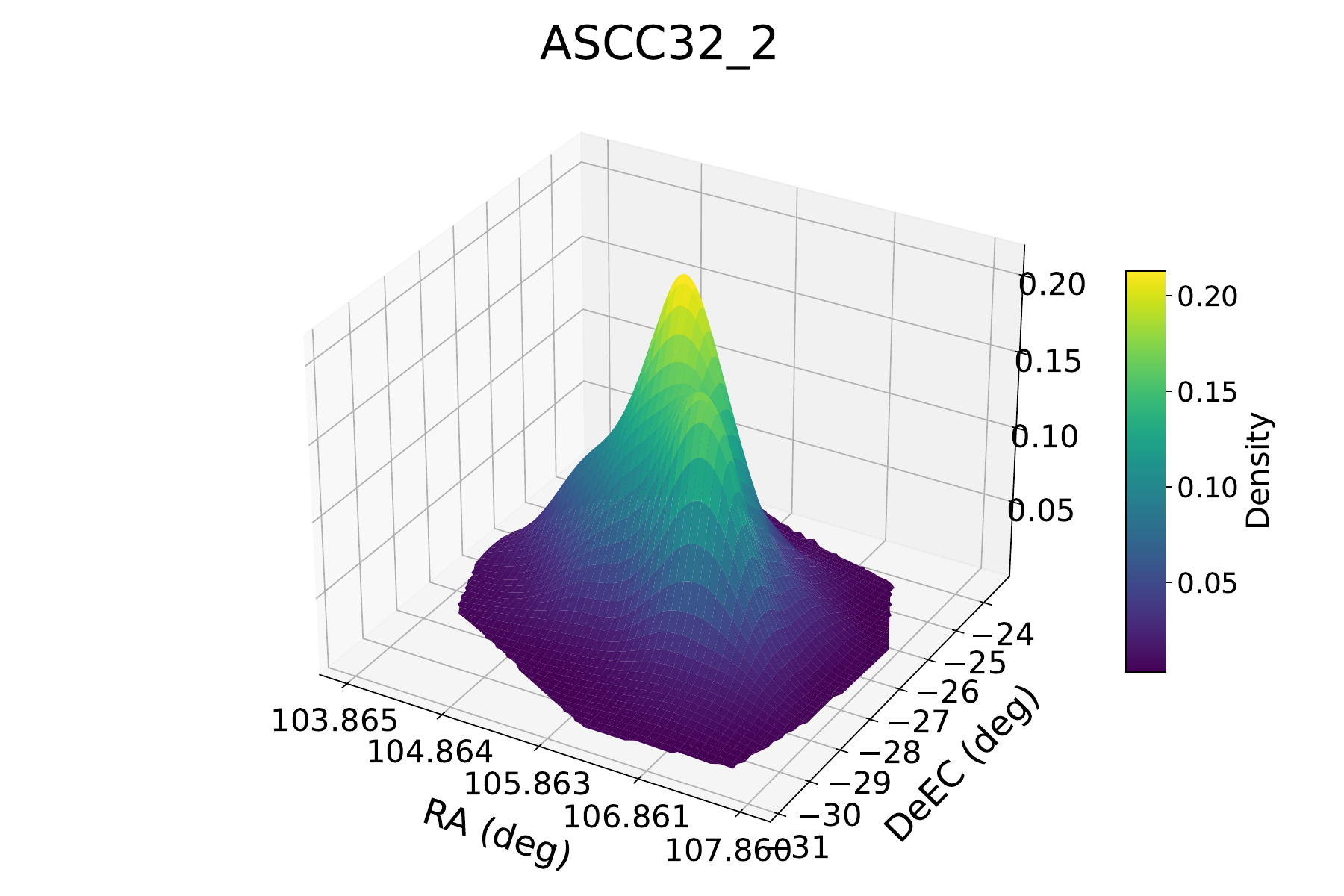}
    
\end{subfigure}
\hfill
\begin{subfigure}{0.49\linewidth}
    \centering
    \includegraphics[width=\linewidth]{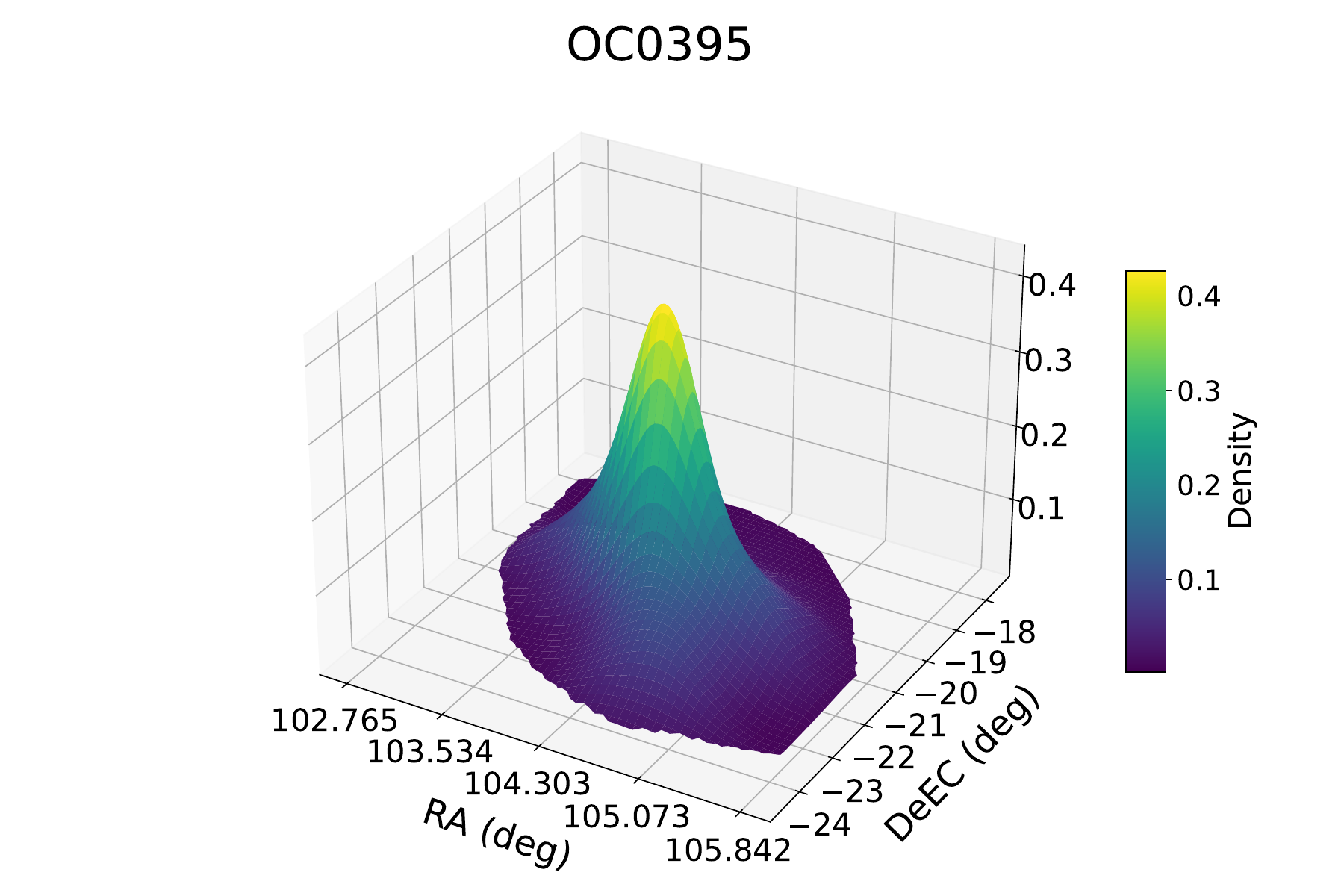}
    
\end{subfigure}
\hfill
\begin{subfigure}{0.49\linewidth}
    \centering
    \includegraphics[width=\linewidth]{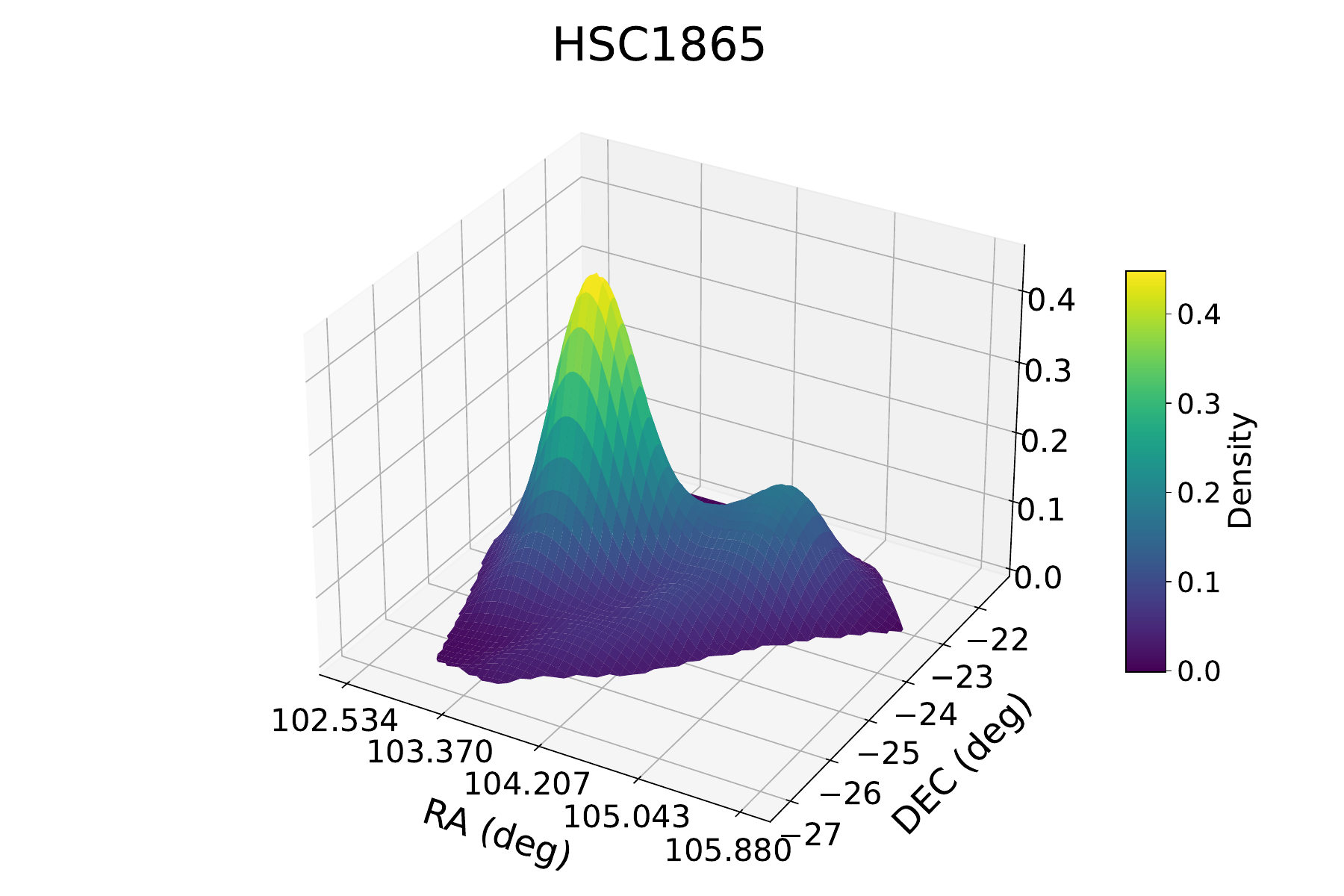}
    
\end{subfigure}

\caption{The Kernel Density Estimate for ASCC\,32$_-$1, ASCC\,32$_-$2, OC\,0395, HSC\,1865. The top-left panel is for the ASCC\,32$_-$1, the top-right panel for ASCC\,32$_-$2, and bottom left panels is for OC\,0395 and bottom right panel is for HSC\,1865. One central dense area for all four parts shows that GMM had separated the Quadruple structure.
\label{3dkde.fig}}

\end{figure*}
\begin{figure*}
\centering
\begin{subfigure}{0.49\linewidth}
    \centering
    \includegraphics[width=\linewidth]{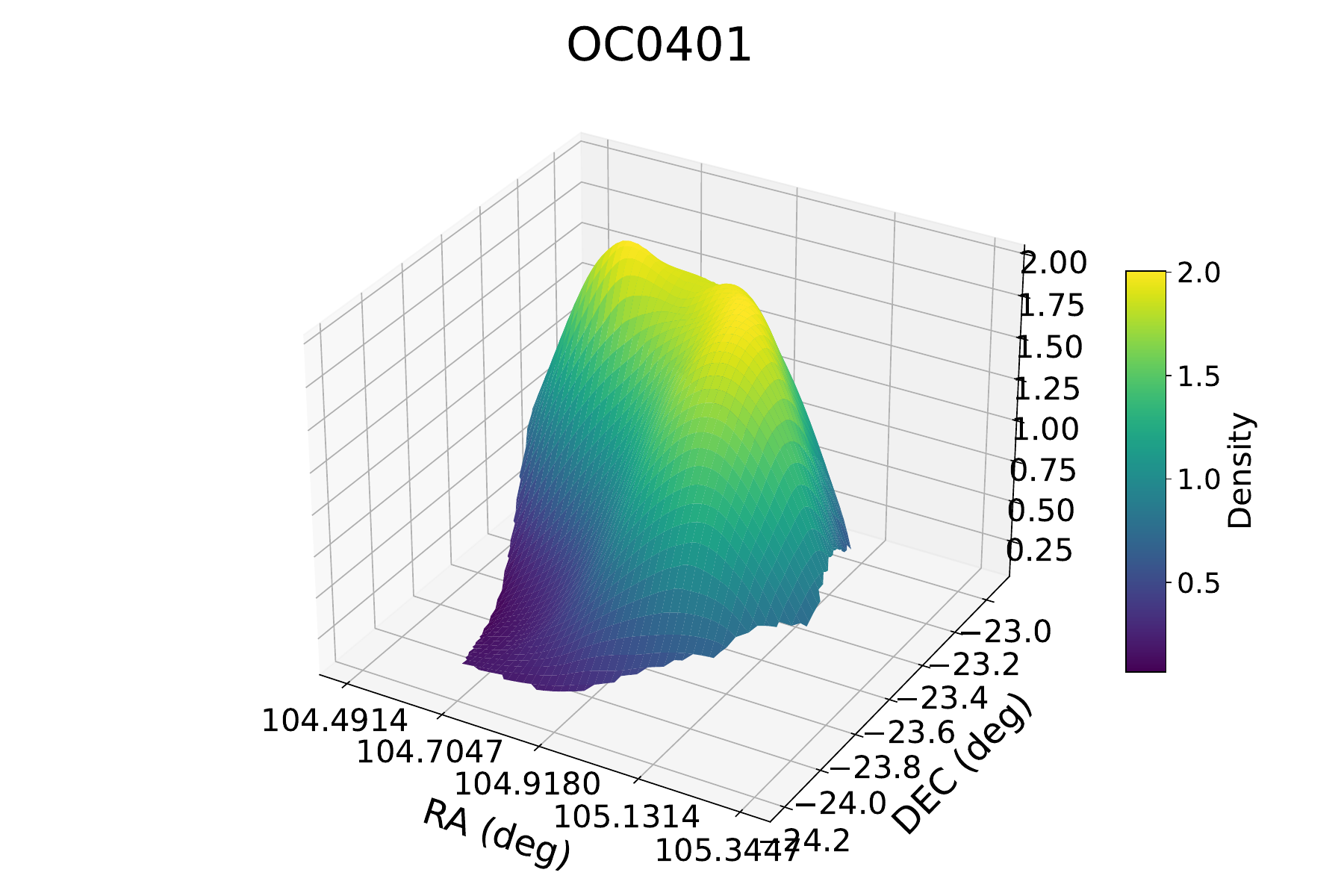}
    
\end{subfigure}
\hfil
\begin{subfigure}{0.49\linewidth}
    \centering
    \includegraphics[width=\linewidth]{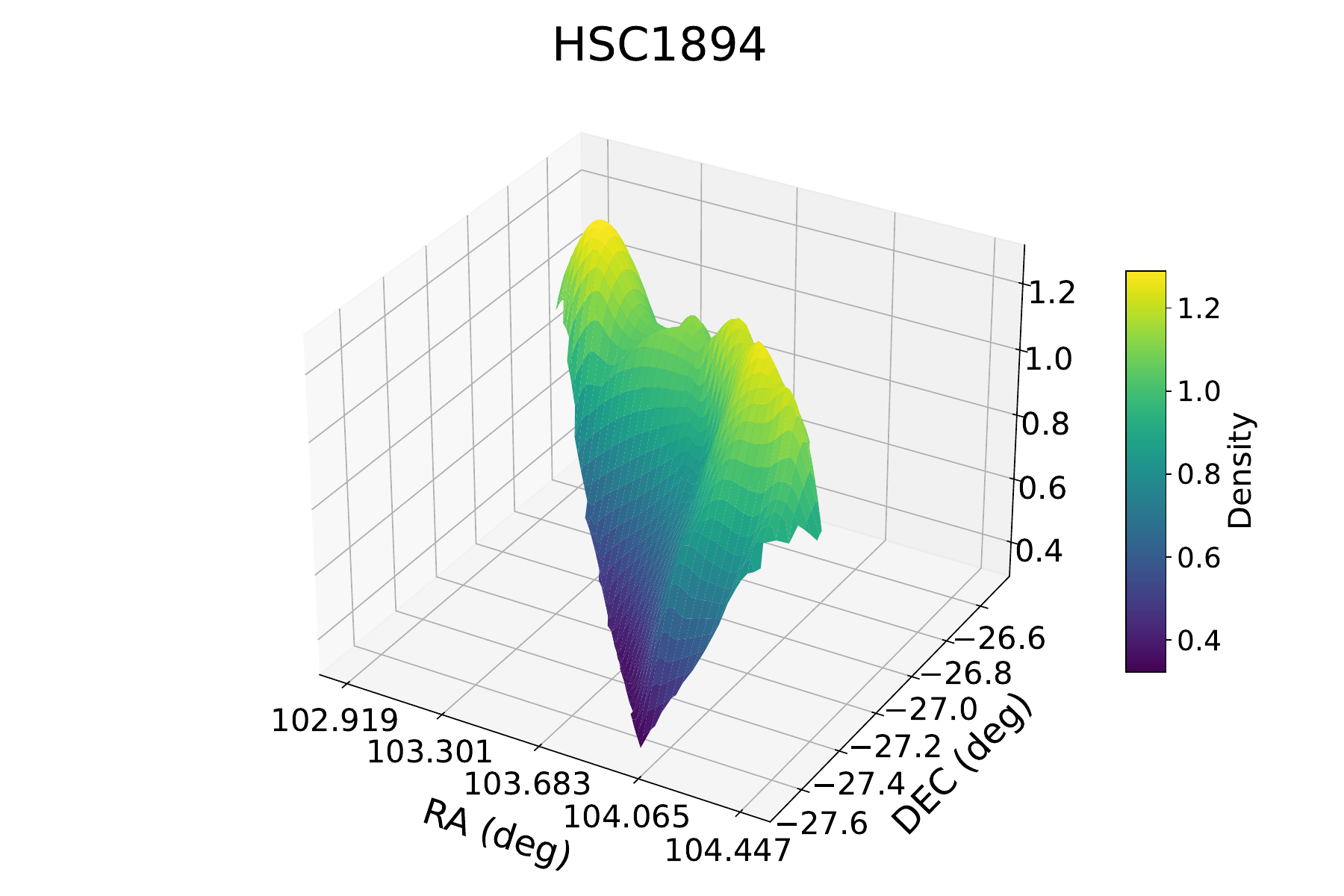}
    
\end{subfigure}
\hfill
\begin{subfigure}{0.49\linewidth}
    \centering
    \includegraphics[width=\linewidth]{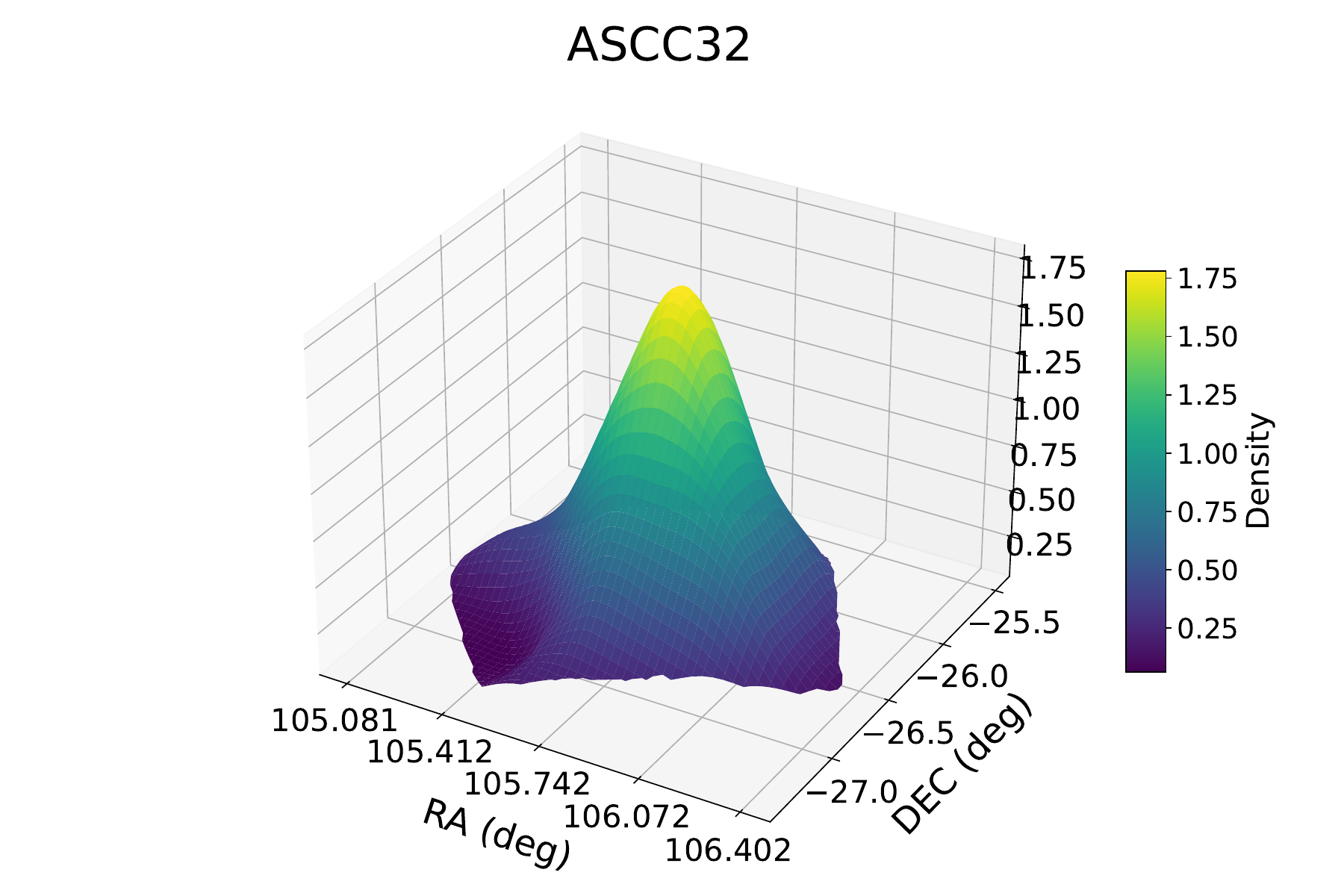}
    
\end{subfigure}
\hfill
\begin{subfigure}{0.49\linewidth}
    \centering
    \includegraphics[width=\linewidth]{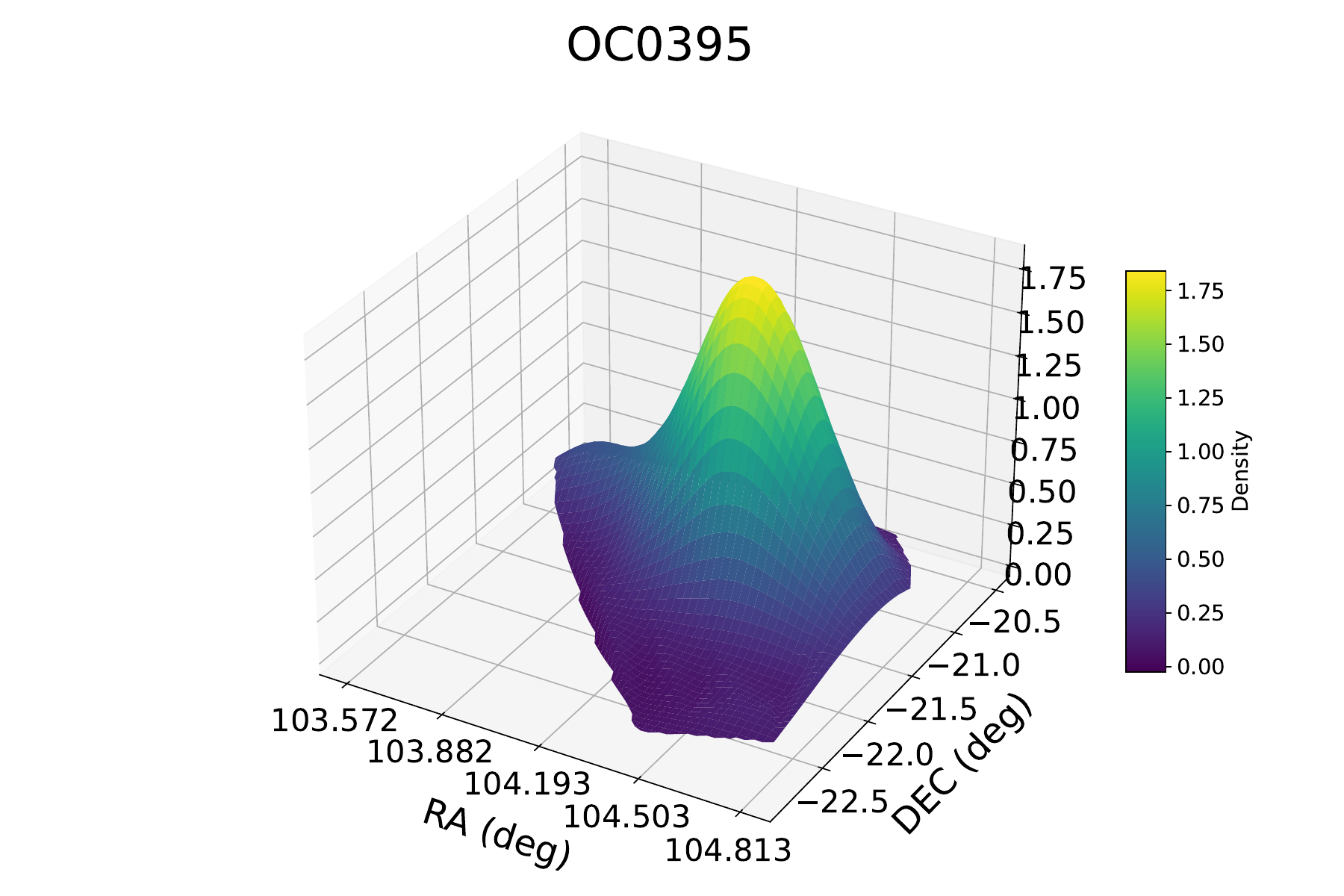}
    
\end{subfigure}
\hfil
\begin{subfigure}{0.49\linewidth}
    \centering
    \includegraphics[width=\linewidth]{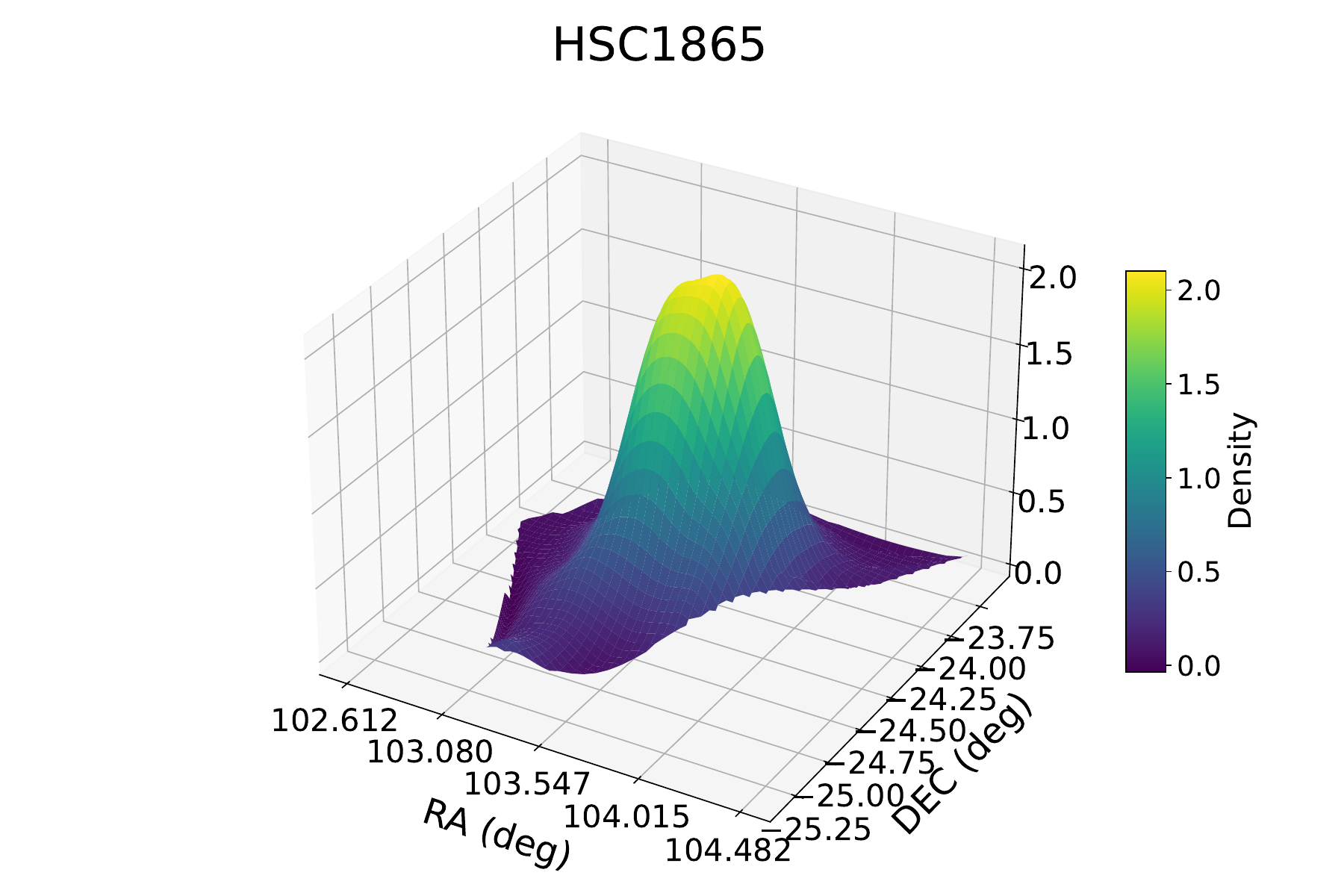}
    
\end{subfigure}
\caption{The Kernel Density Estimate analysis for detecting groups in Hunt and Reffert (2024). The top-left panel for OC\,0401 and top-right panel for HSCC\,1894 do not show central dense region. The middle-left panel for ASCC\,32, The middle-right panel for OC\,0395 and the bottom panel for HSC\,1865 shows one dense region. This shows that only ASCC\,32, OC\,0395 and HSC\,1865 have cluster shapes consistent with the groups detected in our work.
\label{ehtidal.fig}}
\end{figure*}
\begin{table*}
\centering
\caption{Physical parameters of groups in this work. All reported errors are standard errors.}
\begin{tabular}{c c c c c c c c c c}
    \hline
    \hline
    Name & RA~(deg) & DEC~(deg) & Parallax~(mas) & pmRA~(mas\,yr$^{-1}$) & pmDEC~(mas\,yr$^{-1}$) & $v_r$~(km\,s$^{-1}$) & $N_{GMM}$ & D~(pc)\\
    \hline
    ASCC\,32$_-$1 & $105.562\pm0.033$ & $-26.107\pm0.074$ & $1.258\pm0.001$ & $-3.314\pm0.009$ & $3.468\pm0.005$ & 32.365 & 834 & 713.855 \\
    
    ASCC\,32$_-$2 & $105.835\pm0.028$ & $-26.851\pm0.045$ & $1.172\pm0.001$ & $-3.155\pm0.007$ & $3.440\pm0.005$ & 33.730 & 668 & 743.112 \\
    
    OC\,0395 & $104.385\pm0.029$ & $-21.439\pm0.058$ & $1.205\pm0.002$ & $-3.429\pm0.009$ & $3.237\pm0.006$ & 36.924 & 336 & 730.198 \\
    
    HSC\,1865 & $103.973\pm0.025$ & $-24.233\pm0.050$ & $1.162\pm0.002$ & $-2.879\pm0.007$ & $3.477\pm0.005$ & 33.880 & 456 & 743.894 \\
  \end{tabular}
  \label{this work data.tab}
\end{table*}
\begin{table*}
\centering
\caption{Physical parameters of groups in [1]~Kharchenko et al. (2005), [2]~Cantat-Gaudin et al. (2018), [3]~Pang et al. (2022), [4]~Hunt-Reffert (2024), [5]~Alfonso et al. (2024), [6]~He et al. (2022), [7]~Hao et al. (2022). All reported errors are standard errors. }
\begin{tabular}{c c c c c c c}
    \hline
    \hline
    Name & RA~(deg) & DEC~(deg) & Parallax~(mas) & pmRA~(mas\,yr$^{-1}$) & pmDEC~(mas\,yr$^{-1}$) & Number of members\\
    \hline
    
    ASCC\,32~[1] & $105.49$ & $-26.50$ & - & $-3.39\pm0.33$ & $1.97\pm0.35$ & 28 \\
    
    ASCC\,33~[1] & $105.79$ & $-25.05$ & - & $-3.78\pm0.30$ & $3.78\pm0.23$ & 62 \\
    
    ASCC\,32~[2] & $105.714$ & $-26.512$ & $1.230\pm0.004$ & $-3.307\pm0.014$ & $3.477\pm0.008$ & 259 \\
    
    ASCC\,32~[3] &  $105.743\pm0.032$ & $-26.416\pm0.036$ & $1.260\pm0.002$ & $-3.268\pm0.009$ &  $3.478\pm0.006$ & 519 \\
    
    ASCC\,32~[4] & $105.709\pm0.017$ & $-26.365\pm0.024$ & $1.248\pm0.003$ & $-3.217\pm0.007$ & $3.488\pm0.006$ & 194 \\
    
    OC\,0395~[4] & $104.368\pm0.022$ & $-21.353\pm0.032$ & $1.222\pm0.004$ & $-3.468\pm0.009$ & $3.254\pm0.008$ & 143 \\
    
    OC\,0401~[4] & $104.947\pm0.025$ & $-23.526\pm0.035$ & $1.184\pm0.007$ & $-3.069\pm0.019$ & $3.394\pm0.013$ & 69 \\
    
    HSC\,1865~[4] & $103.663\pm0.016$ & $-24.233\pm0.019$ & $1.122\pm0.004$ & $-2.819\pm0.006$ & $3.505\pm0.008$ & 246 \\
    
    HSC\,1894~[4] & $103.267\pm0.069$ & $-26.974\pm0.046$ & $1.218\pm0.005$ & $-2.909\pm0.023$ & $3.603\pm0.015$ & 44 \\
     
    ASCC\,32~[5] & $105.777\pm0.021$ & $-26.676\pm0.030$ & $1.245\pm0.002$ & $-3.229\pm0.009$ & $3.483\pm0.007$ & 432\\
    
    CWNU\,34~[6] &  $106.088\pm0.042$ & $-23.946\pm0.050$ & $1.259\pm0.010$ & $-3.624\pm0.062$ &  $3.384\pm0.017$ & 31	 \\
    
    CWNU\,91~[6] &  $104.338\pm0.022$ & $-21.295\pm0.025$ & $1.230\pm0.005$ & $-3.462\pm0.012$ &  $3.213\pm0.009$ & 218 \\
    
    OC\,0395~[7] & $104.290\pm0.013$ & $-21.272\pm0.021$ & $1.194\pm0.010$ & $-3.486\pm0.048$ & $3.285\pm0.051$ & 73 \\
    
    OC\,0399~[7] & $104.901\pm0.051$ & $-23.149\pm0.012$ & $1.153\pm0.015$ & $-2.932\pm0.085$ & $3.355\pm0.056$ & 21 \\
    
    OC\,0401~[7] & $104.712\pm0.019$ & $-23.362\pm0.013$ & $1.142\pm0.015$ & $-2.988\pm0.065$ & $3.345\pm0.063$ & 26 \\
    
    OC\,0408~[7] & $104.806\pm0.025$ & $-26.449\pm0.016$ & $1.186\pm0.015$ & $-3.257\pm0.110$ & $3.602\pm0.097$ & 20 \\
    
    OC\,0410~[7] & $105.596\pm0.021$ & $-27.313\pm0.018$ & $1.205\pm0.017$ & $-3.090\pm0.038$ & $3.457\pm0.075$ & 34 \\
    
    OC\,0411~[7] & $105.163\pm0.020$ & $-27.643\pm0.017$ & $1.235\pm0.020$ & $-2.910\pm0.144$ & $3.633\pm0.070$ & 18 \\ 
    
  \end{tabular}
  \label{data in other work.tab}
\end{table*}
\noindent
Stars form from primary giant molecular clouds, either as dense clusters or diffuse stellar groups such as OB associations, depending on the cloud's density and turbulence processes~\citep{Lada2003,Maiz-Apell2001,Elmegreen2008apj}. Since stars in clusters and diffuse stellar groups originate from a single parent cloud, they share similar chemical and astrometric properties. These similarities can be observed through both astrometric and photometric data. Stellar associations have weaker gravitational fields than clusters, causing stars to evaporate from them very quickly while, in clusters, stars evaporate over longer periods due to stronger gravitational forces.~\citep{Wright2020}. This duration depends on the number of cluster members, cluster structure, and tidal forces.\\
Several studies have attempted to illustrate the similarities and differences between bound stars in clusters and unbound stellar associations. \cite{Maiz-Apell2001} analyzed the structural properties of massive young clusters and identified a morphological sequence linking super star clusters—with varying halo strengths—to scaled OB associations. This sequence suggests a continuum shaped by the initial mass distribution of their parent molecular clouds. They argued that primary interstellar clouds can form with different structural characteristics: small, dense molecular clouds tend to produce compact clusters, while large, more diffuse molecular clouds give rise to stellar associations.\\ \cite{Elmegreen2008apj} proposed that the pressure within interstellar clouds plays a critical role in determining the nature of stellar group formation. In regions of high pressure, stars tend to form as bound clusters, whereas in low-pressure environments, they are more likely to form as scaled OB associations.\\ 
\cite{H.kirk2011} studied 14 young stellar groups and demonstrated that each exhibits several cluster-like properties, including a broad range of stellar masses, with massive stars tending to form near the center.\\
The field of view of  clusters and OB associations is highly contaminated by field stars because they are located in the inner disk of the Galaxy, making it challenging to detect reliable these group members. Several studies have sought to identify reliable cluster members using statistical methods, such as~\cite{sanders}, based on astronomical parameters. However, as the volume of data increases, the limitations of these methods have become more apparent~\citep{Wu2002,Jilinski2003,Balaguer2004,Javakhishvili2006,Kraus2007,Wiramihardja2009,Krone-Martins2010}.\\ 
In recent years, machine learning algorithms have garnered significant attention among scientists, particularly astrophysicists. Since stars within a cluster and each group of stellar associations share similar astrometric parameters, such as parallax and proper motion, machine learning methods can effectively identify their reliable members, provided that various techniques are employed to enhance the algorithm's accuracy.\\ 
Several studies have explored cluster membership using machine learning approaches based on astrometric and photometric parameters, employing both supervised and unsupervised learning. The most significant of these include: \cite{Gao2013}, \cite{Gao2015}, \cite{ElAziz2016}, \cite{Bhattacharya2017}, \cite{Wilkinson2018}, \cite{Gao2018PASJ}, \cite{Gao2018PASP4c}, \cite{Gao2018m67}, \cite{Cantat-Gaudin2018}, \cite{Castro-Ginard2020}, \cite{Noormohammadi2023}, \cite{Noormohammadi2024}.\\
With the increase in data accuracy and the development of methods capable of handling large datasets, such as machine learning algorithms, some clusters exhibit different structures—binary or multiple—compared to previous studies~\citep{Alejo}.\\
Although several studies have attempted to identify certain stellar systems in our galaxy as binary clusters, further research is required to validate these findings~\citep{Dalessandro,Marcos,Song,Subramaniam1995,Casado}. \cite{Piecka} reviewed a catalog of star clusters and suggested that some are likely binary clusters based on Gaia Data Release 2. Clusters exhibiting identical parameters in the Color Magnitude Diagram~(CMD) and proper motion in Right Ascension~(pmRA) and Declination~(pmDEC), while differing slightly in position and parallax, may be classified as binary systems~\citep{Song}.\\
\cite{Dalessandro} identified several new star complexes near $h$ and $\chi$ Persei using DBSCAN~(Density-Based Spatial Clustering of Applications with Noise). \cite{Piatti} found that some cluster pairs are not true binary systems but rather two nearby clusters, potentially forming a collided system, such as Collinder\,350 and IC\,4665. According to \cite{Marcos}, clusters are considered binary or multiple structures when they share the same age, space velocity, and chemical composition. \cite{2025Hu} studied ASCC\,19 and ASCC\,21, reporting identical proper motion, radial velocity, and CMD distribution for these clusters, leading to the conclusion that they formed as a binary system. \cite{Song} defined binary clusters as those where the difference in proper motion is less than three times the standard deviation~($\sigma$). \cite{Kovaleva2020} demonstrated that Collinder\,135 and UBC\,7 formed as a binary system. Multiple structural configurations are present not only in star clusters but also in OB associations, indicating a diversity of stellar formation environments. For example, \cite{Wilkinson2018} identified two substructures within the Upper Scorpius association based on Gaia Data Release 1, while \cite{Squicciarinimnras2021} increased that number to eight using Gaia Early Data Release 3~(Gaia EDR3), demonstrating the impact of improved data accuracy.\\
ASCC\,32 is located near the Collinder\,132–Gulliver\,21 region, at a distance of 795.83 pc~\citep{Pang2022ApJ}. \cite{Kharchenko2005} provided one of the first reports on ASCC\,32 and ASCC\,33, which are positioned near each other. However, the data from \cite{Kharchenko2005} indicate differences in proper motion in Declination~(pmDEC) between the two clusters, while their apparent distance modulus and proper motion in Right Ascension~(pmRA) are similar, suggesting a possible relationship between them.
\cite{Cantat-Gaudin2018} and \cite{Pang2022ApJ} provided data exclusively for ASCC\,32, which differs in pmDEC compared to \cite{Kharchenko2005} but aligns with the pmDEC values of ASCC\,33 reported by \cite{Kharchenko2005}. \cite{he2022new} reported two newly detected clusters, CWNU\,34 and CWNU\,91, in the ASCC\,32 region. \cite{Alfonso2024} applied HDBSCAN in their study and identified ASCC\,32 based on Gaia Data Release 3~(GDR3). Additionally, \cite{Hao2022A&A} identified six new clusters—OC\,0395, OC\,0399, OC\,401, OC\,0408, OC\,0410, and OC\,0411—within the ASCC\,32 region. \cite{Hunt2024} documented the membership of ASCC\,32, OC\,0401, OC\,0395, HSC\,1894, and HSC\,1865 in this area.\\
\cite{Pang2022ApJ} calculated an age of 25 Myr for ASCC\,32. \cite{Hunt2024} reported ages of 25.051 Myr, 22.42 Myr, and 23.65 Myr for ASCC\,32, OC\,0401, and OC\,0395 respectively, while detecting 11.36 Myr for HSC\,1865 and 36.76 Myr for HSC\,1894, based on 84\% of cluster members. Their results suggest that ASCC\,32, OC\,0401, and OC\,0395 share similar ages.\\
In this work, we investigated the possibility of detecting stellar groups with identical astrometric parameters—proper motion and parallax—within the ASCC\,32 region, as well as a shared main sequence distribution in the Color-Magnitude Diagram~(CMD), while differing in RA and DEC, using machine learning methods. These groups may represent multiple structures, including those that share a common origin and formed simultaneously.\\
To achieve this, we employed DG methods~\citep{Noormohammadi2023}, which integrate two unsupervised machine learning algorithms: DBSCAN and GMM~(Gaussian Mixture Model). In the first step, we applied only proper motion and parallax, and in the subsequent step, we incorporated positional parameters. This approach enables the identification of groups with similar proper motion and parallax using DBSCAN. Ultimately, we expect to observe a unified CMD distribution for these groups, which could indicate a shared origin.\\
The initial motivation for this study stemmed from \cite{Noormohammadi2023}, in which we analyzed several open clusters with varying ages and distances using DG methods based on Gaia EDR~3 within a 150-arcmin field. When ASCC\,32 was examined within this field, DBSCAN detected portions of OC\,0395 that fell within our sample’s field of view. To further investigate, we expanded our field of view to 500 arcmin to encompass both regions, ultimately detecting four structures: OC\,0395, HSC\,1865, and two distinct distributions within ASCC\,32, designated as ASCC\,32$_{-}$1 and ASCC\,32$_{-}$2. These four stellar groups exhibit similar CMDs and closely matching parallax and proper motion values, except for ASCC\,32$_{-}$1, which appears slightly closer than the others. However, they differ in RA and DEC. \\
This study raises an important question: could ASCC\,32$_{-}$1, ASCC\,32$_{-}$2, OC\,0395, and HSC\,1865 have formed as a multiple structure? To explore this possibility, we conducted a detailed investigation of the region, examining astrometric coherence, CMD alignment, and spatial distribution.\\ 
After identifying reliable stellar groups and their respective members, we examined the possible relationships among the studied groups based on their astronomical parameters—parallax, radial velocity, and proper motion—as well as their photometric characteristics, particularly their CMD distributions. These factors may suggest the presence of a multiple structure. We also compared our results with those reported by \cite{Kharchenko2005}, \cite{Cantat-Gaudin2018}, \cite{he2022new}, \cite{Hao2022A&A}, \cite{Pang2022ApJ}, \cite{Hunt2024}, and \cite{Alfonso2024}.\\
In Section~\ref{section.Data}, we describe the conditions for data selection. Section~\ref{section.Method} outlines the methodology. In Section~\ref{result and discussion}, we present the data identified using our method and discuss the findings. Section~\ref{compare} provides a comparison of our results with previous studies. Finally, we summarize our findings in Section~\ref{con}.

\section{Data} \label{section.Data}
\noindent
In this work, our aim is to study stellar groups that share similar astronomical parameters—pmRA, pmDEC, and parallax—within the ASCC\,32 region. \cite{Hunt2024} identified five groups—ASCC\,32, OC\,0395, OC\,0401, HSC\,1865, and HSC\,1894—as having comparable astronomical characteristics. To encompass all these groups, a 500-arcmin radius was selected around the center of ASCC\,32, ensuring the inclusion of most candidate groups members, as detailed in Section~\ref{result and discussion}.\\
We used the latest version of the Gaia data release, Gaia Data Release~3 (GDR3)~\citep{gdr3}. Tables~\ref{uncertainties_astro.tab} and \ref{uncertainties_photo.tab} present the accuracy of GDR~3 in astrometric and photometric parameters based on G magnitude. As shown, for stars fainter than 20 mag in G magnitude, the uncertainties in astrometric parameters increase~\citep{gdr3}.\\
In the initial stages, we needed to determine a distance that would provide a wide field of view of the stellar group morphology, allowing for the detection of the maximum number of group members. \cite{Pang2022ApJ} reported a parallax value in the range of [1.17, 1.37] mas for ASCC\,32, while \cite{Alfonso2024} reported a range of [1.12, 1.41] mas. \cite{Hunt2024} identified ASCC\,32, OC\,0395, OC\,0401, HSC\,1865, and HSC\,1894 within the parallax range of [1.01, 1.38] mas.\\
Within this distance (500-arcmin radius), stars were filtered based on two conditions. For higher accuracy, we selected stars brighter than 20 mag and those with parallax values in the range of [0.9, 1.5] mas, which includes all of the candidate group members, as shown in Section~\ref{result and discussion}, accounting for 528,227 data points. Additionally, selected stars had to have complete data for five astrometric parameters (RA, DEC, pmRA, pmDEC, and Parallax) and two photometric parameters (G magnitude and BP-RP color index). By applying this criterion, the dataset was reduced to 514,754  entries.\\
Finally, the data were normalized using the scale function in the scikit-learn library for preprocessing~(\url{https://scikit-learn.org/stable/modules/generated/sklearn.preprocessing.scale.html}).\\

\section{Method} \label{section.Method}
\noindent
In~\cite{Noormohammadi2023}, we employed a combination of two unsupervised machine learning algorithms—DBSCAN and GMM—to detect reliable open cluster members. DBSCAN identified cluster structures by distinguishing cluster member candidates from field stars, while GMM subsequently classified each cluster using different Gaussian distributions and assigned a probability value to each DBSCAN-detected member. \\
We applied our methods to 12 open clusters varying in age, distance, and number of members. Our approach detected a greater number of cluster members compared to previous studies that utilized machine learning algorithms for cluster membership. Additionally, the DBSCAN-GMM method enabled the identification of diverse cluster structures, including dense and sparse cluster members, as well as nearby and distant clusters within the tidal radius.\\
Since DBSCAN can detect groups of stars that exhibit similar astronomical properties, such as proper motion and parallax within a given region, we applied our previous method (DBSCAN-GMM) to identify stellar group members in the ASCC\,32 region. The methodology is described as follows:
\subsection{DBSCAN}
\noindent
In the initial step, we employed the DBSCAN algorithm to identify a region characterized by stars sharing similar proper motion and parallax, and exhibiting a single main sequence, likely due to a common origin. Several studies have utilized DBSCAN for cluster membership, with some successfully detecting previously unrevealed structures near the cluster region~\citep{Bhattacharya_db, Dalessandro, Prisinzano_yso}.  Also, \cite{Wilkinson2018} used DBSCAN to detect OB association members in Upper Scorpius based on Gaia Data Release~1.\\ 
It requires two parameters: MinPts and Eps. The algorithm considers a circle with a radius equal to the Eps value at the center of each data point. Then, it counts the data points within this circle. If the number of data points in the circle exceeds MinPts, the algorithm considers the central data point a core point. If the number of data points is lower than MinPts but the central data point belongs to the circle with a core point at its center, the algorithm considers this point a border point. Otherwise, the algorithm considers the point as noise.\\ 
The algorithm iterates through all data points, ultimately classifying core and border points as cluster members. One of DBSCAN’s key advantages is its ability to identify diverse cluster distributions within a sample source. This makes it a highly effective algorithm for detecting multiple structures in star clusters.
\subsection{GMM}
\noindent
In the second step, we used the GMM algorithm. The GMM algorithm finds Gaussian distributions in a sample source using the following Equation:
\begin{equation}\label{Eq gaussian.Eq}
  D(G)=\sum_{j=1}^{k}w_{j}(-2\pi)^{-\frac{n}{2}}|\Sigma_{j}|^{-\frac{1}{2}}exp(-\frac{1}{2}(X-\mu_{j})^{T}\Sigma_{i}^{-1}(X-\mu_{j}))
  \end{equation}
\noindent
In Equation~\ref{Eq gaussian.Eq} $D(G)$ is the probability distribution, $X=(x_{1},x_{2},...x_{i})$ is the data points, $n$ is the dimension of the data sets, $\mu_{j}$ and $\Sigma_{j}$ are the mean value and a covariant matrix of the Gaussian distribution, $j$ determines the number of clusters and $w_{j}$ is the weight distribution of the $j^{th} $ Gaussian component.\\
Since cluster stars and some of other stellar groups exhibit a Gaussian distribution in astrometric parameters such as position, proper motion, and parallax, they are well-suited for the GMM algorithm. However, GMM has certain limitations in achieving optimal performance and reliable results, particularly when dealing with large datasets and a low signal-to-noise ratio. To address this, previous studies have filtered data to enhance the signal-to-noise ratio. In our work, DBSCAN accomplished this by utilizing Eps and MinPts.\\

\subsection{DG method~(DBSCAN-GMM method)}
\noindent
Each machine learning algorithm has its own advantages and limitations when analyzing data sources. DBSCAN requires two parameters—Eps and MinPts—and selecting optimal values for these parameters can be challenging. Additionally, GMM relies on a high signal-to-noise ratio for effective performance. By combining these two algorithms, their strengths are enhanced while their weaknesses are mitigated. In our approach, DBSCAN is not used in isolation; instead, MinPts and Eps are carefully chosen to improve the signal-to-noise ratio for GMM in the subsequent step. The combination of DBACAN and GMM is named the DG method.\\
Using the DG method and the maximum likelihood method, \cite{Qiu2024} studied IC\,2488 and IC\,2714, comparing the two approaches. Their findings demonstrated that, for detecting faint stars, the DG method outperformed the maximum likelihood method.\\ 
Selecting hyperparameters for machine learning algorithms is a crucial step in data analysis. If this study relied solely on DBSCAN, we would need to determine reliable values for MinPts and Eps. However, in the DG method, DBSCAN functions primarily as a preprocessing step for GMM in the subsequent stage. Therefore, MinPts and Eps must be chosen to optimize conditions for GMM.\\
In this study, after filtering and normalizing the data as described in Section~\ref{section.Data}, we applied DBSCAN to the ASCC\,32 region using three astrometric parameters: pmRA, pmDEC, and parallax. We then examined the data identified by DBSCAN in the Color-Magnitude Diagram~(CMD) and analyzed the distributions of proper motion and parallax.\\
If indications of group members were present among the field stars—such as the main sequence of stellar group and the proper motion and parallax ranges where group members lie (with a high signal-to-noise ratio)—we applied GMM to these data using five parameters: RA, DEC, pmRA, pmDEC, and parallax. Finally, we analyzed the GMM results in terms of position, proper motion, parallax, and the Color-Magnitude Diagram.\\ 
If DBSCAN provides data that meets the GMM condition—ensuring a high signal-to-noise ratio and an approximate distinction of stellar group structures among field stars, albeit with some contamination—GMM should be able to identify reliable group members. These members appear as a distinct main sequence in the CMD diagram, a well-structured cluster in the position diagram, and a dense region in the proper motion diagram among the sample sources, effectively distinguishing group members from field stars. GMM assigns a probability value to each DBSCAN-detected member. Ultimately, this approach allowed us to identify members with a low contamination rate.\\
With the applied Eps$=0.08$ and an increase in MinPts, the first cluster detected by DBSCAN originates from our sample source. When MinPts is set to 250, two clusters are selected, as shown in Fig~\ref{db_cluster1.fig}. Increasing MinPts to 280, DBSCAN identifies three clusters, which are depicted in Fig~\ref{db_cluster2.fig}. As seen in this figure, the proper motion and parallax diagram indicates that candidate members must belong to C\,3. C\,1 and C\,2 exhibit diverse parallax and proper motion values compared to C\,3 and do not form a localized region in position space. Therefore, we applied the GMM algorithm to the C\,3 cluster.\\
It is clear that a filamentary structure exists within C\,3, to which ASCC\,32 belongs. To extract this structure, we applied the GMM algorithm with a cluster number of 3. Under this configuration, GMM successfully isolated the ASCC\,32 region, which contains several stellar groups. To identify these groups, we applied GMM again using five parameters: RA, DEC, pmRA, pmDEC, and parallax, guided by the Bayesian Information Criterion~(BIC) score. The results are described as follows:\\

\section{Results and Discussions}  \label{result and discussion}
\noindent
DBSCAN identified 8,137 candidate members from the sample source. Fig~\ref{dbkde.pdf} presents the Kernel Density Estimate~(KDE) for DBSCAN-detected members, highlighting three dense regions compared to other areas—one in the ASCC\,32 region and the others in the OC\,0395 and HSC\,1865 regions. All of these were detected in this study, with a higher number of members than in previous works.\\
GMM detected 2,926 stars within the ASCC\,32 region. Fig~\ref{dg_ascc} shows the filament structure identified by GMM, with membership probabilities greater than 0.7. As can be seen, this structure reveals a clear main sequence without any red giant members, suggesting a young stellar population, consistent with findings from previous studies. It also contains several stellar groups, including ASCC\,32, OC\,0395, and HSC\,1865. To separate these groups, we applied GMM based on the Bayesian Information Criterion~(BIC), which indicated the presence of four distinct stellar groups. Fig~\ref{bic.pdf} presents the BIC score used to determine the optimal number of GMM clusters based on RA, DEC, pmRA, pmDEC, and parallax.\\
After applying the GMM algorithm, four group were identified, containing 834, 668, 336, and 456, and 332 members.\\ 
Fig~\ref{ngc_400.fig} presents an astronomical diagram of these groups alongside others, each classified as a separate cluster by the GMM algorithm. In the case of ASCC\,32, we found two distributions located in the same position but with slightly different parallax values, as shown in Fig~\ref{ngc_400.fig}. purple, Blue, yellow, and green dots represent ASCC\,32$_-$1, ASCC\,32$_-$2, OC\,0395, and HSC\,1865, respectively, as referenced in \cite{Hunt2024}. In the next section, we provide a detailed comparison of these results with previous studies.\\
As illustrated in Fig~\ref{ngc_400.fig}, the four groups exhibit distinct peaks in parallax while maintaining closely related structures. They also demonstrate homogeneity in proper motion and the CMD, yet differ in positional structure. A single CMD distribution can indicate a common origin and age.\\
For further details, the radial velocity of each group is shown in Fig~\ref{rv.pdf}. This figure presents a single distribution in radial velocity for ASCC\,32$_-$1,ASCC\,32$_-$2, OC\,0395, and HSC\,1865, providing additional evidence of the relationship between these group as a multiple structure.\\ 
Fig~\ref{ngc3_kde.fig} presents the KDE for ASCC\,32$_-$1, ASCC\,32$_-$2, OC\,0395, and HSC\,1865 in terms of position, proper motion, and parallax. All groups are presented with the same CMD, but ASCC\,32$_{-}$1 exhibits slightly different parallax values and an extended structure in both position and proper motion. The central dense regions of ASCC\,32$_{-}$1 and ASCC\,32$_{-}$2 appear similar in terms of position, although they differ in parallax.\\
As observed, after applying GMM, each component of the multiple structures exhibits a central dense region, indicating that GMM effectively separates individual elements within the system.\\ 
Table~\ref{this work data.tab} presents the astrometric parameters for the four stellar groups detected in this study. As shown in the table, the four groups exhibit coordination in proper motion, distance, and radial velocity, suggesting a possible relationship between them.\\ 
To support the argument that these groups form a multiple structure, three conditions were considered as follows:\\
1. The standard deviation~($\sigma$) for pmRA and pmDEC is [0.271, 0.163], [0.201, 0.143], [0.168, 0.118], and [0.164, 0.120] for ASCC\,32$_-$1, ASCC\,32$_-$2, OC\,0395, and HSC\,1865, respectively. For these clusters, the difference in proper motion in Right Ascension and Declination is less than 3$\sigma$, which aligns with the condition defined by \cite{Song} for binary groups~\citep{Song}.\\
2. As shown in Fig~\ref{ngc_400.fig}, all groups share the same CMD distribution, consistent with the condition defined by \cite{Marcos}.\\
3. Stellar groups must exhibit the same radial velocity distribution~\citep{2025Hu,Song,Marcos}, which is confirmed in Fig~\ref{rv.pdf}.\\
These findings suggest that ASCC\,32$_-$1, ASCC\,32$_-$2, OC\,0395, and HSC\,1865 could form a multiple structure.

\section{Comparison with Other Works}\label{compare}
Table~\ref{data in other work.tab} presents astronomic data for the ASCC\,32 region from various studies. As shown in this table, the parameters of the group represented by blue and purple dots in Fig~\ref{ngc_400.fig} align with ASCC\,32 in \cite{Pang2022ApJ}, \cite{Cantat-Gaudin2018}, \cite{Hunt2024}, \cite{Alfonso2024}, and \cite{Kharchenko2005}, except for pmDEC, which differs in \cite{Kharchenko2005}.\\
In the upper-left panel of Fig.~\ref{compare.fig}, we compare our results with those of \cite{he2022new}, who first reported new groups in the ASCC\,32 region, including CWNU\,34 and CWNU\,91. As illustrated in the figure, OC\,0395 corresponds to CWNU\,91; however, our study improved membership identification and expanded the field of view for this stellar group. Additionally, several members of CWNU\,34 were considered in this work as part of ASCC\,32$_{-}$1 and ASCC\,32$_{-}$2, although CWNU\,34 is not included in either \cite{Hunt2024} or \cite{Hao2022A&A}. This discrepancy may be due to the increased number of identified members for ASCC\,32 in our study.\\ 
The upper-right panel of Fig~\ref{compare.fig} presents a comparison between this study and \cite{Hao2022A&A}. \cite{Hao2022A&A} identified new stellar groups in the ASCC\,32 region, including OC\,0395, OC\,0399, OC\,0401, OC\,0408, OC\,0410, and OC\,0411. In this study, we detected OC\,0395, but as shown in this figure, our results have significantly increased membership. Additionally, OC\,0408, OC\,0410, and OC\,0411 are considered part of ASCC\,32 in our work, as well as in \cite{Hunt2024} and \cite{Alfonso2024}.\\ 
In the bottom-left panel of Fig.~\ref{compare.fig}, we compare the member detection results for ASCC\,32 from this study with those of \cite{Alfonso2024}, who applied HDBSCAN based on Gaia DR~3. Our study identified a greater number of members and expanded the field of view for ASCC\,32, allowing us to distinguish two substructures: ASCC\,32$_{-}$1 and ASCC\,32$_{-}$2.\\
The bottom-right panel of Fig.~\ref{compare.fig} presents five stellar groups reported by \cite{Hunt2024}—ASCC\,32, OC\,0395, OC\,0401, HSC\,1865, and HSC\,1894—in the ASCC\,32 region based on Gaia DR~3. As shown in Fig.~\ref{compare.fig}, ASCC\,32, OC\,0395, and HSC\,1865 correspond to four groups identified in this study. Additionally, HSC\,1894 and OC\,0401 were detected as parts of other groups in this study using the GMM algorithm.\\
Detection of members within a 500-arcmin radius from ASCC\,32 revealed a relationship between ASCC\,32$_{-}$1, ASCC\,32$_{-}$2, OC\,0395, and HSC\,1865 as a multiple structure. The DG method identified more members for these four components compared to other studies reviewed in this work. In this study, OC\,0401 from \cite{Hunt2024} is located in the dense region between ASCC\,32 and OC\,0395. It may have formed due to gravitational interactions between these groups, warranting further investigation.\\
Fig~\ref{pmc_co.fig} (left panel) compares the proper motion distributions of ASCC\,32$_{-}$1 and ASCC\,32$_{-}$2 in this work with those presented in \cite{Alfonso2024}. It is evident that our study detected two distinct proper motion distributions, whereas both were considered a single distribution in \cite{Alfonso2024}. The right panel compares our results for ASCC\,32$_{-}$1 and HSC\,1865 with OC\,0401. In this study, the proper motion of OC\,0401 is identified as overlapping with ASCC\,32$_{-}$1 and HSC\,1865. Based on its proper motion distribution, OC\,0401 does not exhibit a cluster-like structure.\\
For further investigation, we calculated the 3D-KDE for groups selected in this work and in \cite{Hunt2024}. Fig~\ref{3dkde.fig} presents the 3D distribution of the KDE in position for ASCC\,32$_{-}$1, ASCC\,32$_{-}$2, OC\,0395, and HSC\,1865 in this study. All groups exhibit a single peak in the position distribution, which is a characteristic property of open clusters.\\
Fig.~\ref{ehtidal.fig} shows the KDE distributions for ASCC\,32, OC\,0395, OC\,0401, HSC\,1894, and HSC\,1865. As illustrated, the KDE analysis of the ASCC\,32 region—used to detect stellar groups in \cite{Hunt2024}—indicates that only ASCC\,32, OC\,0395, and HSC\,1865 exhibit a single dense region, consistent with the stellar groups identified in this study. In contrast, OC\,0401 and HSC\,1894 do not show this property. Consequently, the GMM algorithm did not identify a Gaussian distribution in RA and DEC for OC\,0401 and HSC\,1894.\\
This study provides a broader view of the ASCC\,32 region's morphology and reveals that all groups detected around ASCC\,32 in \cite{he2022new}, \cite{Hao2022A&A}, and \cite{Hunt2024} collectively form a four-part structure.\\

\section{Conclusions} \label{con}
\noindent
Research on binary and multiple structures in the Milky Way has advanced with increasing data accuracy and the development of powerful analytical methods. The third edition of the Gaia data release offers an excellent opportunity to explore various aspects of astrophysics. Machine learning algorithms have identified data with similar properties across different sources. With the support of two facilities, we can further investigate multiple structures within our galaxy.\\
In this study, we identified multiple structures within 500 arcmin around the ASCC\,32 region using a combination of two unsupervised algorithms (DBSCAN-GMM). This analysis relied solely on astronomical parameters and included only stars brighter than 20 mag, as data uncertainty increases for fainter stars.\\
We detected two distributions for ASCC\,32, which we refer to as ASCC\,32$_{-}$1 and ASCC\,32$_{-}$2 in this work. We also identified members of OC\,0395 and HSC\,1865. These objects exhibit similar parameters in pmRA, pmDEC, and parallax, as well as a consistent main sequence structure in the color-magnitude diagram. These findings align with data from other studies on these groups, suggesting that ASCC\,32$_{-}$1, ASCC\,32$_{-}$2, OC\,0395, and HSC\,1865 may have originated from the same interstellar cloud. Additionally, the difference in proper motion among these clusters is less than three times the standard deviation, indicating that they are moving together.\\
We separated the four groups using the GMM algorithm and demonstrated that, after applying GMM, the KDE for each group revealed a single dense region, illustrating that GMM effectively distinguishes each component. We identified a greater number of members for ASCC\,32, OC\,0395, and HSC\,1865 compared to previous studies. Additionally, our analysis indicates that OC\,0401 and HSC\,1894 are substructures of HSC\,1865 and ASCC\,32 rather than distinct groups.\\
Moving forward, the structure and formation of multiple clusters should receive greater attention through simulation methods, with comparisons between simulation results and observational data.\\

\section*{DATA AVAILABILITY}
The data used in this work are Gaia DR3~\citep{gdr3} available at \url{https://gea.esac.esa.int/archive/} and we are ready to send our data to any research request.

\bibliographystyle{mnras}
\bibliography{ref}

\end{document}